\documentclass[aps,twocolumn,preprintnumbers,prb,amsmath,amssymb,amsfonts, superscriptaddress,floatfix]{revtex4}

\usepackage{amsfonts}
\usepackage{amssymb}
\usepackage{amsmath}
\usepackage[normalem]{ulem}
\usepackage{color}
\usepackage{xcolor}
\usepackage{latexsym}
\usepackage{graphicx}
\usepackage{subfigure}
\usepackage{graphics}
\usepackage{floatflt}
\usepackage{epsfig}
\usepackage{overpic}
\usepackage{soul} 
\usepackage{varioref,xr-hyper} 
\usepackage[utf8]{inputenc}
\usepackage{mathbbol}
\usepackage{makecell}

\newcommand{\be}{\begin{equation}}
\newcommand{\ee}{\end{equation}}
\newcommand{\bea}{\begin{eqnarray}}
\newcommand{\eea}{\end{eqnarray}}

\def\lb{\label}

\newcount\bozza \bozza=0
\ifnum\bozza=1
\newdimen\shift \shift=-1.0truecm
\def\lb#1{%
{\label{#1}\rlap{\kern\shift{$\scriptstyle#1$}}}}
\else\def\lb#1{\label{#1}} \fi

\begin{document}

\title{Sense-Making phase transition framework for Affect}

\author{Miguel Bosch}
\thanks{These authors contributed equally to this work.}
\affiliation{Facultad de Psicología, Universidad Complutense de Madrid,Campus de Somosaguas, E-28223 Pozuelo de Alarcón, Madrid, Spain}

\author{Laura Fanfarillo}
\thanks{These authors contributed equally to this work.}
\affiliation{Istituto dei Sistemi Complessi (ISC-CNR), Via dei Taurini 19, I-00185 Rome, Italy}
\author{Gustavo Diez}
\affiliation{Nirakara Lab, Institute of Research and Cognitive Science, Madrid, Spain}
\author{Victor G\'omez Mayordomo}
\affiliation{Department of Neurology, Institute of Neuroscience, Hospital Universitario Vithas Madrid La Milagrosa, Vithas Hospital Group}

\author{J. Ricardo Arias-Gonzalez}
\affiliation{Nanophotonics Technology Center, Universitat Polit\`ecnica de Val\`encia, Camino de Vera s/n, 46022 Valencia, Spain}
\author{Bel\'en Valenzuela}
\affiliation{Instituto de Ciencia de Materiales de Madrid, 
ICMM-CSIC, Cantoblanco, E-28049 Madrid, Spain}

\date{\today}
 
\begin{abstract}
Contemporary accounts of affect emphasize different perspectives, ranging from the regulation of bodily states in allostatic approaches to the relational and sense-making character stressed by enactive theories. Here, we show that affect can be made explicit within the sense-making phase transition framework previously introduced for the perception of an inner sensation. Within this same formal structure, each stable minimum of the sense-making potential defines an affective state: its position determines valence, while its depth determines arousal. Affect therefore is not introduced as an independent ingredient, but emerges as the associated state in the valence--arousal space, providing a link between allostatic and enactive approaches to affect. This formulation distinguishes valence intensity from arousal, allows for dominant and secondary affective states, and naturally accommodates mixed-valence configurations beyond a strictly bipolar description. Revisiting representative landscapes of the model, we show how neutral, uncertain, polarized, and mixed-valence regimes arise as a function of contextual sensory information, learned historical evaluation, and externally supplied information. We further show that the resulting valence--arousal relation provides a theoretical basis for the upper-open V-shaped organization discussed in the affect literature, while also opening a natural route toward a dynamical description of affect and emotions.
\end{abstract}

\maketitle

\section{Introduction}
\label{sec:intro}

Affect plays a central role in human experience, shaping how individuals evaluate sensations, regulate behavior, and adapt to changing internal and external conditions\cite{Panksepp1998, Pfaff2005}. It is also deeply implicated in health, well-being, decision making, and mood disorders. Yet, despite its importance, affective science still lacks a fully agreed descriptive and theoretical framework. Valence and arousal are widely regarded as its fundamental dimensions\cite{Russell1980, BarrettRussell99, Posner:2005aa}, but their precise definition and mutual relation remain contested.

Recent accounts of affect have emerged from different perspectives. One important line of work, developed by L.F.~Barrett, discusses affect as an ongoing evaluative state involved in allostasis, that is, in the regulation of bodily resources under changing conditions\cite{Barrett:2009aa, LFBarrettBook, Satpute19Barrett}. This evaluation process is closely related to interoception, understood broadly as the sensing and interpretation of bodily states. In this view, affect is not a detached mental episode, but part of the organism's continuous adaptive regulation. In parallel, enactive approaches have developed the notion of sense-making as the process through which organisms establish meaningful relations with their world\cite{Varela:2016aa,Di-Paolo:2017aa}. This perspective becomes directly relevant to affect in the work of G.~Colombetti, who argues that cognition is inherently affective because sense-making is intrinsically evaluative, and that affect is best understood as embodied appraisal rather than as a pre-given property of experience\cite{ColombettiThompson2007,Colombetti2014,Colombetti2015}.

An important difference remains in how these approaches account for affective experience. Barrett emphasizes the role of bodily regulation and interoceptive processes\cite{Barrett:2015aa}, whereas the enactive approach understands affectivity as intrinsic to sense-making, grounded in the organism's autonomous and normatively organized engagement with its environment. This organization involves interrelated domains of normativity --organic, sensorimotor, social, and existential-- which can mutually enable, constrain, and sometimes conflict with one another. Within this enactive framework, mental disorders may involve disruptions in the coordination of different normative domains, including patterns of excessive rigidity or instability that constrain the organism's adaptive capacities and its ability to maintain autonomous, flexible engagement with its environment\cite{Di-Paolo:2017aa, Garcia:2025aa}. Barrett's account likewise emphasizes that mood disorders arise when affective patterns become rigid and persistent in an evolving socio-environmental context\cite{LFBarrettBook, Greenwood:2025aa}.
Thus, although these approaches arise from different perspectives and assign explanatory priority to different levels of organization, they nevertheless converge in viewing affect as an emergent form of ongoing appraisal, and in highlighting the persistence of affective patterns, despite changing internal and external conditions, as a central feature of psychopathology.

In this context a central challenge arises: to develop a formal account of affect in which valence and arousal emerge naturally from a coherent mathematical description of evaluation, while remaining compatible with both the allostatic perspective and the enactive notion of sense-making. Such a framework should also provide a principled basis for understanding how disruptions or rigidifications of evaluative organization may give rise to affective disorders. In this work, we argue that the sense-making phase transition framework recently developed for the perception of an inner sensation\cite{Valenzuela_FrontPain2024,Fanfarillo_RSOS2026}, and rooted in condensed matter field theory, provides such a formal basis for affect at the level of evaluative experience. 

The paper is organized as follows. In Sec.~\ref{sec:overview},  we briefly review those aspects of affect research that are most relevant to the present work and discuss how recent allostatic and enactive perspectives motivate the sense-making approach.
In Sec.~\ref{sec:model}, we recall the Landau sense-making framework and its relevant statistical-mechanical ingredients. In Sec.~\ref{sec:affect}, we introduce the affective interpretation of the sense-making landscape through representative examples, showing how valence and arousal are associated with its minima and how different affective configurations can be identified. In Sec.~\ref{sec:landscapes}, we analyze in more detail how these affective regimes evolve as the sensory context is varied, following the emergence and reorganization of affect within the same representative landscape configurations. In Sec.~\ref{sec:valence_arousal}, we derive the resulting relation between valence and arousal and show how symmetric and asymmetric upper-open V-shaped organizations arise within the framework. Finally, we discuss the conceptual implications of the approach in Sec.~\ref{sec:discussion}, briefly outline its temporal extension toward a theory of dynamical affect and emotions in Sec.~\ref{sec:outlook}, and summarize our main results in the Conclusions.

\section{Affect, valence, and arousal: overview and sense-making perspective}
\label{sec:overview}

Valence and arousal have been extensively discussed in affect research. For broader historical and conceptual accounts we refer the reader, for example, to Refs.~[\onlinecite{Barrett:2009aa, Satpute19Barrett, ColombettiKuppens2024}] Here we only recall a few milestones that are particularly relevant for the present work and for motivating the sense-making perspective adopted below. 

\begin{figure}[b]
\centering
\includegraphics[width=0.7\linewidth]{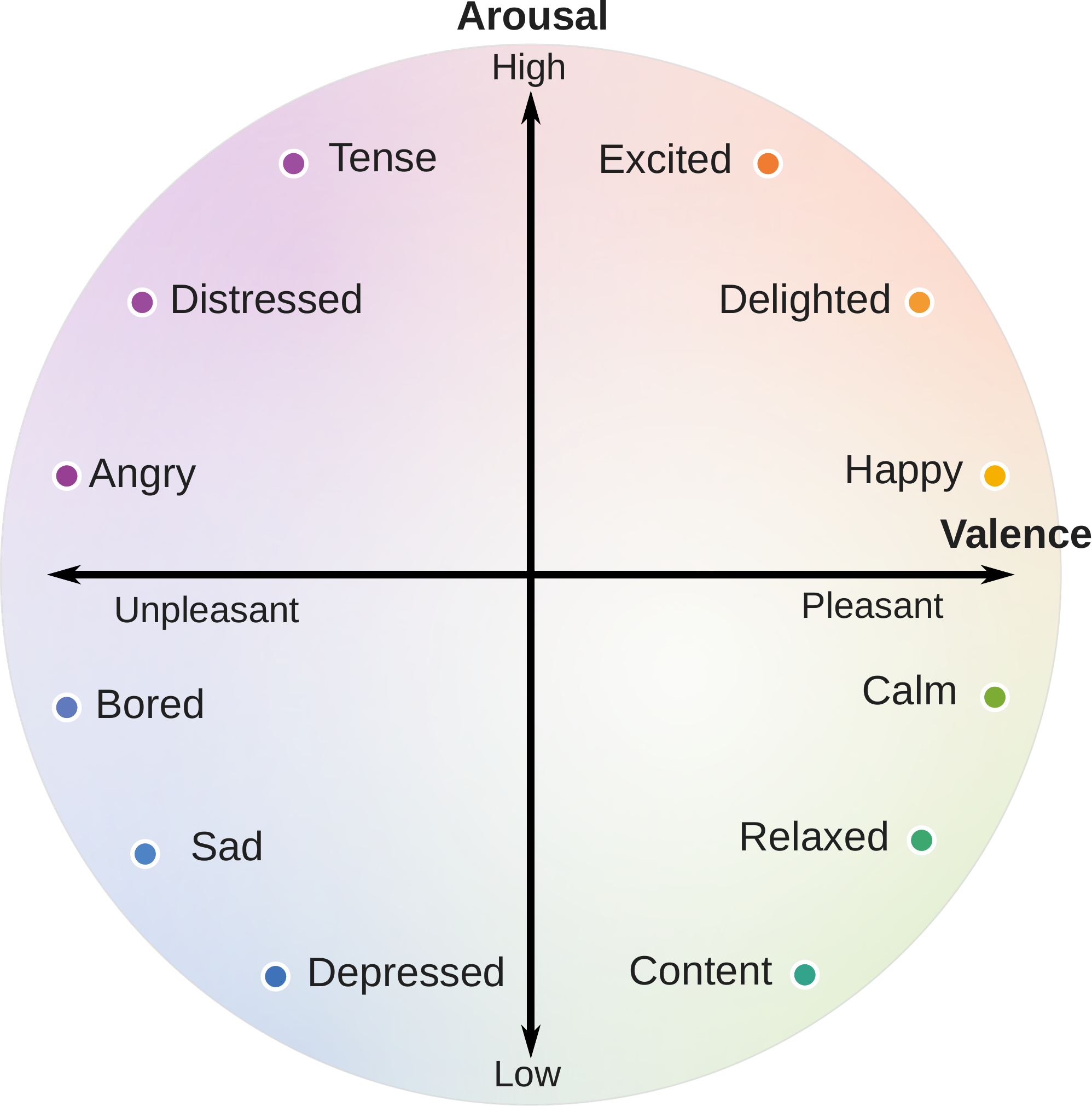}
\caption{Circumplex representation of affective states. The different states are arranged in the valence--arousal plane according to their position along the pleasant--unpleasant dimension and their level of arousal. We emphasize that this representation is simply a descriptive visual organization of affective experience, not the result of a model based on an underlying theoretical principle.}
\label{fig_circumplex}
\end{figure}

\subsection{Brief overview and motivation}
W. Wundt was among the first researchers to propose that affect involves two properties: valence, which refers to how a sensation is evaluated as pleasant or unpleasant, and arousal, which ranges between quiet and active\cite{Wundt1924}. A classical visual organization of affective states in terms of these two dimensions is the circumplex model proposed by J.~Russell, schematically shown in Fig.~\ref{fig_circumplex}. In this representation, affective states are arranged in a valence--arousal plane, and arise from cognitive interpretations of sensations \cite{Russell1980, BarrettRussell99, Posner:2005aa}.

\begin{figure}[t]
\centering
\includegraphics[width=0.92\linewidth]{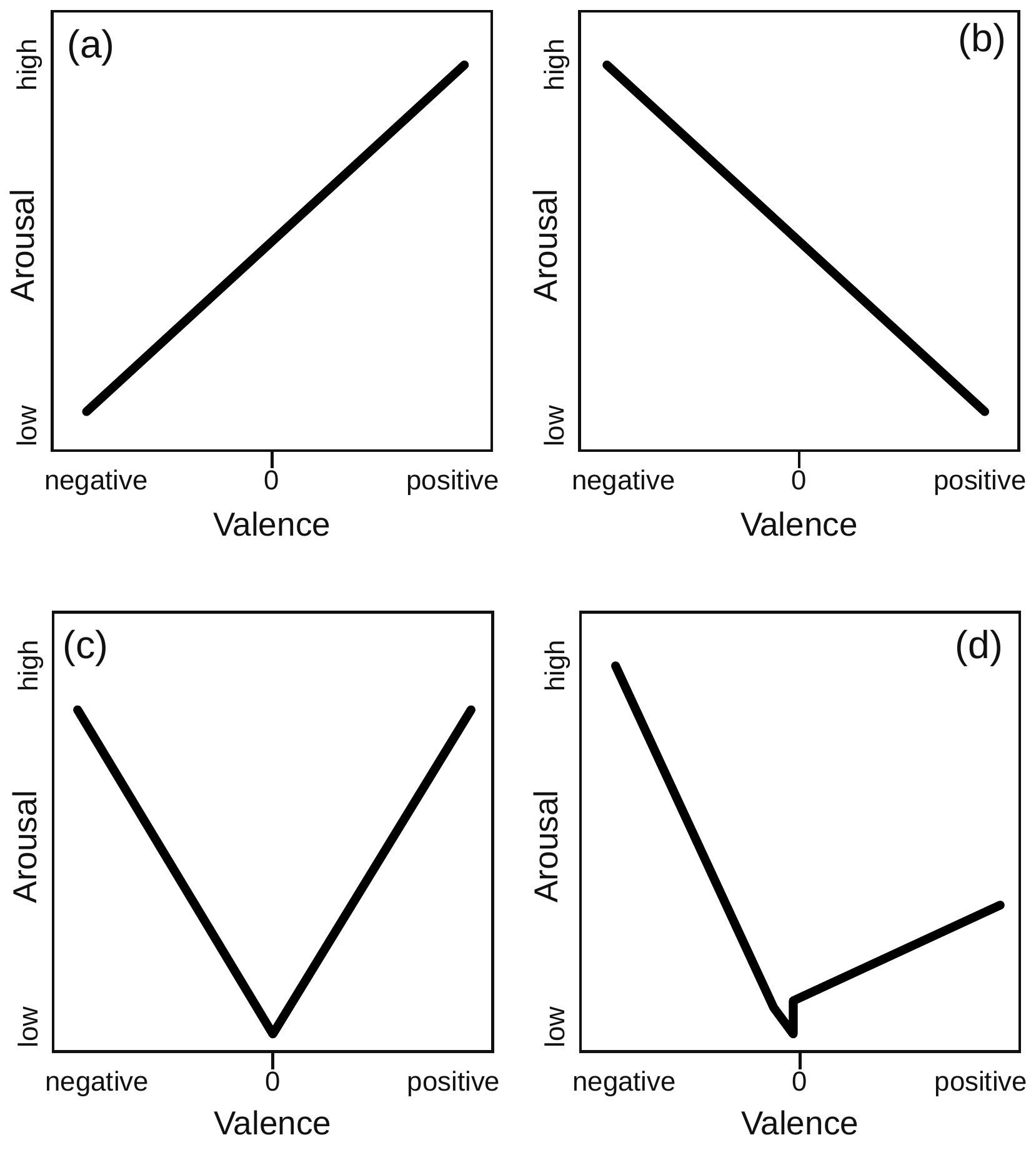}
\caption{Representative proposals for the relation between valence and arousal, as reviewed by P.~Kuppens et al.~[\onlinecite{Kuppens:2013aa}]: (a) positive linear relation, (b) negative linear relation, (c) symmetric upper-open V-shaped relation, and (d) asymmetric upper-open V-shaped relation. As emphasized in Ref.[\onlinecite{ColombettiKuppens2024}], these schematics should not be regarded as equally grounded in empirical evidence or as necessarily describing directly comparable phenomena, since valence and arousal are used with different meanings across the literature.}
\label{fig_V}
\end{figure}

Since then, the precise definition of valence and arousal, as well as the relation between them, has remained a central question in affect research. The circumplex picture has been increasingly questioned, including by its original authors, particularly in clinical contexts where valence and arousal do not appear to behave as independent dimensions\cite{Lang1992Motivational}. 
As reviewed by P.~Kuppens et al.~[\onlinecite{Kuppens:2013aa}], several different proposals about the valence--arousal relation have been formulated in the literature. We schematically report a few representative examples in Fig.~\ref{fig_V}. 
This discussion is further complicated by the fact that both {\it valence} and {\it arousal} are used in the literature in multiple senses that are not always clearly distinguished. Valence, for example, may refer to felt pleasantness or unpleasantness, the positive or negative character of an emotion, approach--withdrawal tendencies, or properties attributed to a stimulus or appraisal\cite{ColombettiKuppens2024}. Arousal is similarly non-unitary, encompassing subjective activation, bodily or autonomic mobilization, wakefulness, and engagement with evocative stimuli\cite{Satpute19Barrett}. Different models may therefore use the same terminology while referring to partly different phenomena.

At the empirical level, psychological assessments and physiological measures often suggest a relation between valence intensity (how pleasant or unpleasant a sensation feels) and arousal (how calm or activated it feels), while also indicating that they should be treated as distinct aspects of affective experience\cite{Kuppens:2013aa}. The valence--arousal relation is often found not to be flat or strictly independent, but to organize along an upper-open V-shaped, often asymmetric, pattern, as the one shown in Fig.~\ref{fig_V}(d), in which arousal tends to increase as valence becomes more intense\cite{Lang1993LookingPictures, Lang1998Emotion, Bradley1999Measuring, Bradley2001Emotion, Ito2005Variations, Kuppens:2013aa}.

It has also been suggested that valence is not necessarily organized along a single bipolar dimension (pleasant {\it vs} unpleasant), but may instead reflect the relative contribution of two orthogonal dimensions, positive affect and negative affect, thereby accommodating mixed affective states\cite{Watson1988, Russell2017}. 
Within this line of work, the Affective Ising Model was recently proposed as a computational model of affect dynamics in an orthogonal positive- and negative-affect space \cite{Loossens:2020aa}, capable of describing non-Gaussian empirical features of affective data, including asymmetry and mixed states. The model successfully reproduces some empirical affective patterns, but, by construction, it does not address how affective organization itself, together with the corresponding valence--arousal properties, emerges from the organism's evaluative relation to its environment.

These developments clarify both the progress already made and the open problem in current affect research. On the one hand, the literature has identified recurrent structures in affective experience, supported by physiological measures, including asymmetric valence--arousal relations and the possibility of mixed states. On the other hand, the available descriptions and formal models do not yet explain how this affective organization arises from the organism's ongoing evaluative relation to a sensation within a meaningful context. It is precisely at this point that the sense-making perspective becomes relevant for the experienced affect.

\begin{figure}[tbh]
\includegraphics[width=0.9\linewidth] {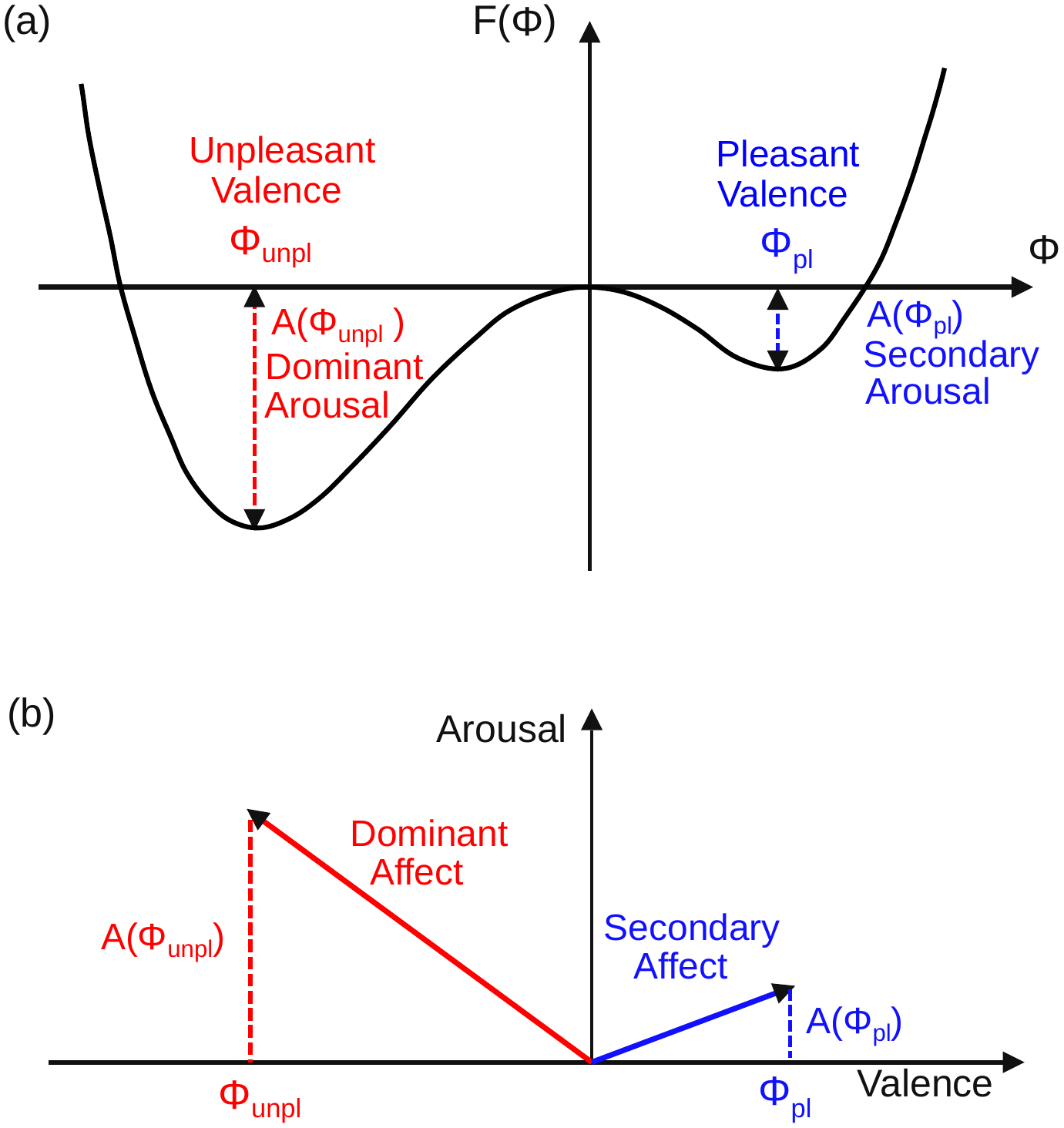}
\caption{Schematic construction of affect from the sense-making framework. (a) $F(\phi)$ is the sense-making landscape. The positions of the local minima, $\Phi_{\rm unpl}<0$ and $\Phi_{\rm pl}>0$, define unpleasant (in red) and pleasant valence (in blue), respectively, while the depth of each minimum relative to the reference level $F(\phi)=0$ defines its arousal. The deeper minimum identifies the dominant affective state, whereas the shallower minimum corresponds to a secondary state. In the example shown, the dominant affect has unpleasant valence and larger arousal, while the secondary affect has pleasant valence and smaller arousal. (b) Visual representation of the affect as a vector in the valence--arousal plane using the same color code.}
\label{fig_affect_landscape}
\end{figure}

\subsection{A sense-making framework for affect } 
In the sense-making framework\cite{Valenzuela_FrontPain2024, Fanfarillo_RSOS2026} the evaluation of a sensation is described by an order parameter and a sense-making effective potential whose coefficients encode contextual sensory information, learned historical valuation, and socially mediated bias. The resulting landscape selects the stabilized evaluative stances toward the sensation. We review the formal construction in the next section. What matters here is that this same landscape already contains the ingredients needed to make affect explicit: the position of the minimum defines valence, with its sign determining whether the experience is pleasant or unpleasant and its magnitude quantifying its intensity, while its depth provides a natural measure of arousal, as shown in Fig.~\ref{fig_affect_landscape}(a). Affect can therefore be formulated within the existing sense-making framework, without modifying its formal structure.

In this way, the sense-making framework might provide a formal bridge between affect as ongoing evaluative organism regulation and affect as embodied sense-making. It shows that valence and arousal are distinct but generally not independent quantities, because both emerge from the same landscape that organizes the evaluation of the sensation. 
This construction naturally allows for mixed affective states whenever locally stable pleasant and unpleasant states coexist within the same landscape, without introducing positive and negative affect as separate components of the model. When these states have different depths, the framework also distinguishes dominant and secondary affective states, as shown in Fig.~3(b).
Moreover, by following how the minima evolve across contexts, the framework provides a natural basis for understanding the emergence of different valence--arousal relations discussed in the affect literature. The aim of the present paper is to establish the static foundations of this construction. We focus on how valence, arousal, and affect can be defined within the sense-making framework, while only briefly indicating the minimal time-dependent extension from which a dynamical account of affect and emotions can later be developed.

\section{Landau model for the valence of a sensation}
\label{sec:model}

\begin{table*}[tbh]
\begin{center}
\begin{tabular}{|c|c|}
\hline
Symbol & Definition \\ \hline
$\phi$ &
\parbox[c]{11.8cm}{\centering \rule{0pt}{3.2ex} Order parameter defining the valence: $\phi<0$ unpleasant, $\phi>0$ pleasant, $\phi=0$ neutral.\rule[-1.4ex]{0pt}{0pt}} \\ \hline
$F(\phi)$ &
\parbox[c]{11.8cm}{\centering \rule{0pt}{3.2ex}Sense-making potential determining the relative stability of valence states.\rule[-1.4ex]{0pt}{0pt}} \\ \hline
$h_{\text{ext}}$ &
\parbox[c]{11.8cm}{\centering \rule{0pt}{3.2ex} Social polarization: external bias favoring one valence sector.\rule[-1.4ex]{0pt}{0pt}} \\ \hline
$h_{\text{int}}$ &
\parbox[c]{11.8cm}{\centering \rule{0pt}{3.2ex}Learned historical valuation: internal history-dependent bias skewing the potential.\rule[-1.4ex]{0pt}{0pt}} \\ \hline
$a=a_0(T-T_0)$ &
\parbox[c]{11.8cm}{\centering \rule{0pt}{3.2ex}Quadratic coefficient controlling neutral-state stability. \rule[-1.4ex]{0pt}{0pt}} \\ \hline
$T$ &
\parbox[c]{11.8cm}{\centering \rule{0pt}{3.2ex} Sensory context: high $T$ noisy context, low $T$ informative context about the sensation.\rule[-1.4ex]{0pt}{0pt}} \\ \hline
$T_0$ &
\parbox[c]{11.8cm}{\centering \rule{0pt}{3.2ex} Contextual threshold at which the neutral state loses local stability for $h_{\text{ext}}=0$.\rule[-1.4ex]{0pt}{0pt}} \\ \hline
$a_0$ &
\parbox[c]{11.8cm}{\centering \rule{0pt}{3.2ex}Intrinsic prefactor setting the $T$-variation of the quadratic term $a(T)$ around $T_0$.\rule[-1.4ex]{0pt}{0pt}} \\ \hline
$\kappa$ &
\parbox[c]{11.8cm}{\centering \rule{0pt}{3.2ex} Stiffness controlling the cost of non-uniform $\phi$ across the lattice of vital norms.\rule[-1.4ex]{0pt}{0pt}} \\ \hline
$b$ &
\parbox[c]{11.8cm}{\centering \rule{0pt}{3.2ex}Intrinsic quartic coefficient stabilizing the potential and shaping the curvature near the minima.\rule[-1.4ex]{0pt}{0pt}} \\ \hline
\end{tabular}
\caption{Definitions of the quantities entering the sense-making framework for affect.}
\label{tab:definitions}
\end{center}
\end{table*}

In this section, we recall the sense-making phase transition framework developed in Refs.~[\onlinecite{Valenzuela_FrontPain2024, Fanfarillo_RSOS2026}], emphasizing the ingredients relevant for the discussion of affect.

As in effective field theories of many-body systems\cite{Altland:2010aa, Fetter_2003, Fradkin_2021}, the framework does not aim to explicitly represent all degrees of freedom simultaneously involved in the organism's experience. Rather, one identifies the collective variable relevant to the phenomenon under study and constructs an effective description for that particular sense-making process. Here we consider a single sensation and retain one scalar field, $\phi$, describing its evaluative organization, from which we construct the sense-making potential $F(\phi)$ associated with this sensation. Other concurrent sensations or cognitive processes are not assumed to be absent; insofar as they modify how informative the sensory context is with respect to the focal sensation, their influence enters through the contextual variable $T$. If two or more sensations are themselves the object of interest and their mutual interaction is relevant, the framework can instead be generalized to a multichannel description involving coupled fields. This {\it question-dependent selection} of the collective degrees of freedom retained explicitly follows the standard logic of effective field theory\cite{Altland:2010aa, Fetter_2003, Fradkin_2021}: depending on the phenomenon of interest, one may construct a theory for a single field or for several coupled fields. 

\subsection{Microscopic basis of the sense-making potential} 
A microscopic derivation of the sense-making potential was developed in Ref.~[\onlinecite{Fanfarillo_RSOS2026}]. For a selected sensation, the microscopic model is a spin model defined on a conceptual network of interacting vital norms. Each site of the network represents a specific vital norm which may reflect interrelated organic, sensorimotor, social, and existential levels of normativity. 
This is analogous to other abstract mental spaces used in the literature characterized by distributions of beliefs\cite{Friston13} or networks of beliefs, behaviors and feelings\cite{AttitudeModel2016}.
In our model, the spin variable associated with that site encodes the evaluation assigned to the norm in relation to the selected sensation and can take the values $+1$, $-1$, or $0$, denoting respectively positive, negative, or neutral evaluations. For example, in the case of hunger, relevant norms or judgments may concern {\it avoiding excessive weight gain}, {\it the body's need for energy replenishment after physical exercise}, or {\it the possible influence of weather changes} on the sensation. For a given individual, these norms may be evaluated respectively as negative, positive, or neutral in relation to the hunger sensation, corresponding to spin values $-1$, $+1$, and $0$. A coarse-graining treatment of these interacting evaluations introduces the continuous field $\phi(x)$, which represents their local collective organization, and leads to the Landau sense-making potential $F(\phi)$ used below.
The coordinate $x$ therefore refers to the coarse-grained evaluative space defined by the network of vital norms, rather than to physical space. Within the microscopic spin-model formulation\cite{Fanfarillo_RSOS2026}, the Landau coefficients can be expressed explicitly in terms of microscopic quantities. 

\subsection{Sense-making potential and adapted states}
At the coarse-grained level, the more general form of the sense-making potential is \cite{Altland:2010aa}:
\begin{eqnarray}
F(\phi) &=& \int d^d x   
\Bigg[ -h_{\text{ext}}\,\phi(x)
+\frac{a}{2}\phi(x)^2
+\frac{\kappa}{2}\bigl(\nabla\phi(x)\bigr)^2 \nonumber \\
& \quad & \quad \quad \quad
+\frac{h_{\text{int}}}{3}\phi(x)^3
+\frac{b}{4}\phi(x)^4 \Bigg].
\label{eq:model_general}
\end{eqnarray}
Here $\phi(x)$ is the scalar order parameter, encoding the local valence of the inner sensation: $\phi(x)<0$ corresponds to unpleasant valence, $\phi(x)>0$ to pleasant valence, and $\phi(x)=0$ to a neutral state. The Landau coefficients $h_{\text{ext}}$, $a$, $b$, $h_{\text{int}}$ represent the intrinsic individual parameters and psychosocial factors that modulate the sense-making. The coordinate $x$ spans the coarse-grained evaluative space defined by the network of vital norms introduced above. The sense-making potential results from the integration over the whole available space $\int d^d x$. The gradient term, weighted by $\kappa$, accounts for the cost of non-uniformity in the evaluative field across this space. $\kappa$ controls the coherence of the evaluative field: 
large $\kappa$ suppresses local variations and favors alignment across the entire evaluative space of vital norms, whereas small $\kappa$ permits local differentiation. In this regard, $\kappa$ may be interpreted as a structural proxy  of mental rigidity and perceptual or affective coherence, modulating the balance between global alignment and local flexibility in evaluative dynamics\cite{Fanfarillo_RSOS2026}. The significance of the coefficients of Eq.~\eqref{eq:model_general} is summarized in Table I. 

In the present work we focus on the homogeneous limit, in which the order parameter is uniform across the lattice, $\phi(x)=\phi$. In this case the gradient term vanishes by definition, and one recovers the simplified form discussed in Ref.~[\onlinecite{Valenzuela_FrontPain2024}]:
\begin{equation}
F(\phi) = -h_{\text{ext}}\phi + \frac{a}{2}\phi^2 + \frac{h_{\text{int}}}{3}\phi^3 + \frac{b}{4}\phi^4.
\label{eq:model}
\end{equation}
The designation {\it sense-making potential} reflects its statistical-mechanics role: $F(\phi)$ assigns relative probability, and hence stability, to alternative macroscopic evaluations of the sensation,
\begin{equation}
P(\phi)\propto e^{-F(\phi)}.
\end{equation}
The minimization of $F(\phi)$ is therefore a stability principle: local minima identify the most likely and stable valence states. 
The global minimum of the landscape, where automatic attention is placed, defines the state selected by the organism. Within the sense-making framework, we refer to it as the \textit{adapted state} to emphasize that it emerges from the relation between the organism and its sensed internal and external world. Importantly, ``adapted'' does not imply convergence toward an externally defined optimum: the position of the global minimum depends on both contextual influences and the organism's intrinsic and historically acquired parameters. Adaptation is therefore relational, individual, and history-dependent. This usage is consistent with recent enactive approaches to mental health, which move away from a strict adaptive/maladaptive dichotomy, and emphasize a plural and graded notion of normativity grounded in the organization and history of the organism\cite{Garcia:2025aa}.

\subsection{Sensory context and Landau coefficients} 
The shape of the sense-making landscape is determined by the specific value of the Landau coefficients and of the temperature. 

The temperature $T$ parametrizes the contextual sensory evidence, namely how informative the overall context is with respect to the specific sensation under study. High $T$ corresponds to a noisy or weakly informative background, that favors, in the absence of an external field $h_{\text{ext}}$, a neutral, non-polarized evaluation
, whereas low $T$ corresponds to a context that is largely informative about the sensation and allows the stabilization of a finite pleasant/unpleasant evaluation.

The coefficients of the sense-making potential of Eq.(\ref{eq:model}) encode distinct sources of influence on the evaluation of the sensation. Their definition in terms of the microscopic parameters of the spin model can be found in Ref.[\onlinecite{Fanfarillo_RSOS2026}]. Here, we briefly recall their role in shaping the sense-making landscape and their cognitive meaning.

As in Landau theory\cite{Altland:2010aa} the quadratic coefficient $a$ explicitly depends on temperature as 
\begin{equation}
a(T)=a_0(T-T_0).
\label{eq:adef}
\end{equation}
$T_0$ is the threshold at which the neutral evaluation $\bar\phi=0$ loses local stability in the absence of an external bias $h_{\text ext}$ and $a_0>0$ is an intrinsic individual parameter that controls how fast the system approaches the instability. 

The quartic coefficient $b$ is an intrinsic individual parameter that shapes the curvature of the functional around the minima, and to stabilize the overall potential, it needs to be positive $b>0$.

$h_{\text{ext}}$ and $h_{\text{int}}$ are the {\it external} and the {\it internal} bias respectively. The external bias can be associated to socially mediated cues or expert information that directly polarize the macroscopic valence state. The internal bias is history-dependent and accumulated through prior experience associated with the sensation. 
In the absence of both biases, the sense-making potential is symmetric between positive and negative evaluations, i.e. $F(\phi)=F(-\phi)$. A finite $h_{\text{int}}$ or $h_{\text{ext}}$ breaks this symmetry, making the two valence sectors inequivalent. With the convention of Eq.~\eqref{eq:model}, $h_{\rm ext}>0$ ($h_{\rm ext}<0$) favors pleasant (unpleasant) valence, whereas $h_{\rm int}>0$ ($h_{\rm int}<0$) favors unpleasant (pleasant) valence. Their effects on the landscape are nevertheless distinct: for $h_{\text{ext}}=0$, a finite $h_{\text{int}}$ preserves the neutral stationary state $\bar{\phi}=0$ at all $T$, while progressively favoring one of the two polarized sectors as $T$ is lowered, i.e., as the sensory context becomes more informative.
A finite $h_{\text{ext}}$, instead, directly shifts the stationary state away from neutrality, so that the high-$T$ adapted state itself acquires a finite valence. Notice that the distinction between external and internal bias has a direct empirical counterpart in pain: socially conveyed and conditioned expectations about pain modulate pain reports through dissociable brain networks\cite{koban-2019, wager-2015}, consistent with treating $h_{\rm ext}$ and $h_{\rm int}$ as separate terms of the landscape.


%
%

\section{Affect from the Landau model}
\label{sec:affect}

As illustrated in Fig.~\ref{fig_affect_landscape} the sense-making framework allows a straightforward definition of the affect and of its dimension -- valence and arousal -- as two distinct but not independent quantities:
\begin{itemize}
\item[-]{\bf {Valence}} 
For a given $F(\phi)$, the stationary con\-fi\-gu\-ra\-tions are defined 
by the extrema of the sense-making potential and the adapted state is selected as the global minimum among them. 
The associated value of the order parameter, $\bar{\phi}$, defines the valence state: its sign determines the pleasant or unpleasant character, with $\bar{\phi}>0$ corresponding to pleasant valence and $\bar{\phi}<0$ to unpleasant valence, while its magnitude $|\bar{\phi}|$ measures  the valence intensity. The neutral state corresponds to $\bar{\phi}=0$.
\item[-]{\bf {Arousal}} Arousal is associated with the depth of the corresponding minimum. Since, for the functional in Eq.~\eqref{eq:model}, the neutral reference satisfies $F(0)=0$, the arousal associated with a minimum at $\bar{\phi}$ is quantified as
\begin{equation}
A(\bar{\phi})=-F(\bar{\phi}).
\end{equation}
Arousal therefore emerges from the landscape itself and measures the stability and relevance of the corresponding state.
\item[-]{\bf {Affect}} As illustrated in Fig.~\ref{fig_affect_landscape}(b) affect is defined as the corresponding state in the valence--arousal space,
\begin{equation}
\text{Aff} (\bar{\phi}) = \bigl(\bar{\phi},A(\bar{\phi})\bigr).
\end{equation}
A natural consequence of this definition is the possibility of mixed affective states, whenever the same landscape supports locally stable affective states in both valence sectors.
\end{itemize}

\begin{figure*}[tbh]
\includegraphics[width=0.97\textwidth] {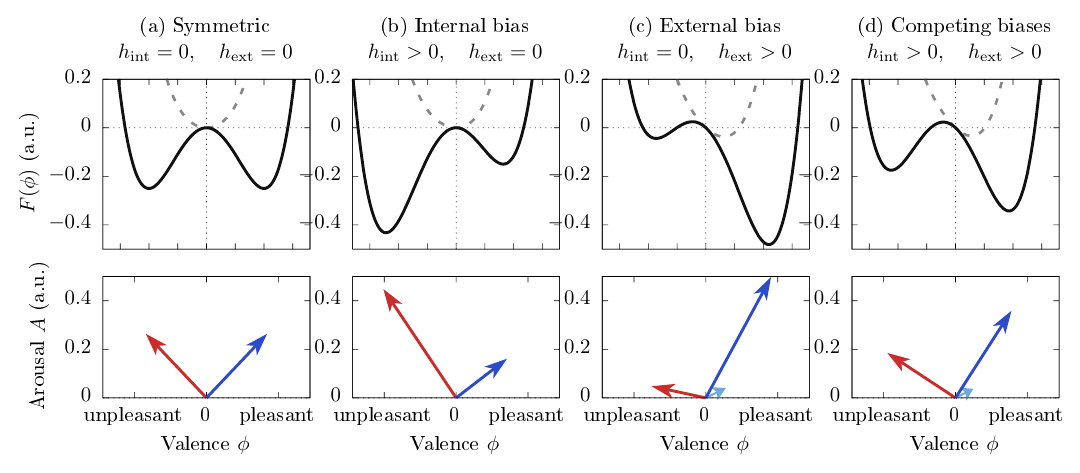}
\caption{Representative affective organizations generated by the sense-making landscape. Top row: high- (dashed grey line) and low-$T$ (solid black line) landscapes for (a) the symmetric case, (b) finite internal bias, (c) finite external bias, and (d) competing internal and external biases; all other model parameters and temperatures are kept fixed across panels. Bottom row: corresponding affective states in the valence--arousal plane, with valence given by the position of each minimum, $\bar{\phi}$, and arousal by its depth, $A=-F(\bar{\phi})$. The symmetric case yields equivalent pleasant and unpleasant states, while finite biases generate dominant and secondary affects; an external bias also produces finite valence and arousal already at high $T$. When both biases are finite, they may reinforce the same valence sector or compete by favoring opposite sectors; panel (d) illustrates the competing case. The color code is as in Fig.~\ref{fig_affect_landscape}: red arrow for unpleasant and blue arrow for pleasant. The light blue arrow indicates that a finite affect is present already at high-$T$.}
\label{fig:affect_examples}
\end{figure*}

To make this affective construction explicit, we now analyze a few representative cases illustrated in Fig.~\ref{fig:affect_examples}. In each case, we compare the sense-making landscape at high and low $T$ and identify the corresponding affective states from the position and depth of the stable minima.

\paragraph*{{\bf Symmetric affective landscape}} As we already mentioned, if we consider $h_{\text{ext}}=h_{\text{int}}=0$, the sense-making potential is symmetric between the two valence sectors, $F(\phi)=F(-\phi)$. As shown in Fig.~\ref{fig:affect_examples}(a) the neutral solution, $\bar{\phi}_0$, is the only stationary state at high $T$. Below the critical threshold $T_0$, the neutral state loses stability and two equivalent minima emerge: $\phi_{pl}$ and $\phi_{unpl}$, associated, respectively, with pleasant and unpleasant valence states. The two polarized states have equal magnitude and correspond to degenerate minima of the potential, so that, by definition, they carry the same arousal:
\begin{equation}
A=-F(\phi_{pl})=-F(\phi_{unpl}) .
\end{equation}
The affective states associated with the pleasant and unpleasant states are, respectively
\begin{equation}
\text{Aff}_{pl} = (\phi_{pl},A), \qquad \text{Aff}_{unpl} = (\phi_{unpl},A).
\end{equation}
and can be represented as vectors in the valence-arousal plane as shown in the corresponding panel of Fig.~\ref{fig:affect_examples}This provides the simplest example of the mixed affective configuration introduced above: two locally stable affective states of opposite valence coexist, here with equal arousal.
As in the standard Landau description of spontaneous symmetry breaking, the symmetric landscape does not privilege either valence sector; below $T_0$, a particular realization spontaneously selects one of the two equivalent affective states, which then defines the adapted state.

At this stage, it is useful to introduce two bias measures. The valence bias is
\begin{equation}
\Delta \phi = |\phi_{pl}|-|\phi_{unpl}|,
\end{equation}
while the potential bias is
\begin{equation}
\Delta F = |F(\phi_{pl})|-|F(\phi_{unpl})|.
\end{equation}
These quantities characterize two distinct asymmetries of the affective landscape: $\Delta\phi$ measures the imbalance in valence intensity, i.e. in the distance of the two states from neutrality, whereas $\Delta F$ measures the imbalance in their stabilization and, consequently, in their arousal. In the symmetric case, both vanish, $\Delta\phi=0$ and $\Delta F=0$, reflecting the exact balance between the two polarized branches. 

\paragraph*{{\bf Affective bias}} Both the internal bias $h_{\rm int}$ and the external bias $h_{\rm ext}$ can lift the symmetry between the two valence sectors, although they modify the sense-making landscape in different ways. As we will discuss in the next section, the temperature evolution of the sense-making landscape is richer than in the symmetric limit. Here we only focus on the qualitative change between the high-$T$ regime and the broken-symmetry state at low $T$.

Let us consider first the effect of the internal bias alone, setting $h_{\rm ext}=0$. We take, for example, $h_{\rm int}>0$, as shown in Fig.~\ref{fig:affect_examples}(b). 
While at high $T$ the adapted state remains neutral, as in the unbiased case, at low $T$ the landscape clearly shows that the unpleasant minimum is global and the pleasant one is only a local minimum. Since the global minimum is deeper than the local one, the two states are associated with different arousal values,
\begin{equation}
A_{\rm unpl}=-F(\phi_{\rm unpl}), \qquad
A_{\rm pl}=-F(\phi_{\rm pl}),
\end{equation}
with $A_{\rm unpl}>A_{\rm pl}$. The larger arousal associated with the unpleasant valence identifies it as the dominant affective state,
\begin{equation}
\text{Aff}_{\rm dom}=\text{Aff}_{\rm unpl}
=(\phi_{\rm unpl},A_{\rm unpl}),
\end{equation}
whereas the pleasant state defines the secondary affect,
\begin{equation}
\text{Aff}_{\rm sec}=\text{Aff}_{\rm pl}
=(\phi_{\rm pl},A_{\rm pl}).
\end{equation}
The emergence of a dominant and a secondary affective state is conceptually important. 
The internal bias lifts the degeneracy of the mixed affective configuration, so that pleasant and unpleasant affective states can coexist within the same sense-making landscape with different stability and arousal.
The vector representation of the asymmetric mixed-state configuration is shown in the corresponding affective panel of Fig.~\ref{fig:affect_examples}(b). 

This asymmetry is captured quantitatively by the valence and potential bias introduced above. In the present case, both are finite. A nonzero $\Delta\phi$ indicates that the two valence states no longer have the same intensity, while a nonzero $\Delta F$ indicates that they are no longer equally stabilized and therefore carry different arousal. 
The larger $|\Delta F|$, the more strongly the landscape privileges one affective possibility over the other: the dominant state becomes more stable and more aroused than the secondary one. We explicitly discussed the case $h_{\text{int}}>0$. An opposite bias does not change the phenomenology discussed here, but simply reverses the configuration, with the pleasant state becoming the dominant affect and the unpleasant one the secondary affect.

Considering instead $h_{\rm int}=0$ and a finite external bias $h_{\text{ext}}$, a qualitatively different behavior emerges already in the high-$T$ regime. In contrast to the case of a finite internal bias, for which the high-$T$ stationary state remains neutral, $h_{\text{ext}}\neq0$ shifts the stationary state away from $\phi=0$, producing an affective state with finite valence and finite arousal. At lower $T$, when two minima of opposite valence coexist, the external bias makes them inequivalent, leading again to a dominant and a secondary affective state and hence to a mixed-affect configuration, characterized by finite valence and potential biases $\Delta\phi$ and $\Delta F$. The sign of $h_{\text{ext}}$ determines which valence sector is favored. In Fig.~\ref{fig:affect_examples}(c) we show the high- and low-$T$ landscapes and the corresponding affective state for a case $h_{\text{ext}}>0$ favoring the positive valence sector, notice the light blue arrow indicating that a finite affect is present already at high-$T$.  

Finally, when both $h_{\rm int}$ and $h_{\rm ext}$ are finite, their effects combine in determining the resulting affective organization. The two biases may favor the same valence sector, reinforcing the asymmetry of the landscape, or opposite sectors, partially compensating one another. The latter case is shown in Fig.~\ref{fig:affect_examples}(d). The resulting dominant and secondary affective states therefore depend on the relative direction and strength of the learned evaluation encoded by $h_{\rm int}$ and the current external information represented by $h_{\rm ext}$. In this sense, the same external cue can have different affective consequences depending on the pre-existing internal bias on which it acts.

\paragraph*{{\bf Implications for affective organization}} These examples illustrate the main conceptual consequences of the construction. First, affect is generated directly by the sense-making landscape, rather than introduced as an additional variable. 
Second, it distinguishes valence intensity from arousal: the former is measured by $|\bar{\phi}|$, whereas the latter is determined by the depth of the corresponding minimum.
Third, the framework naturally accommodates mixed-valence configurations, since the same sensation may support coexisting pleasant and unpleasant affective states within the same sense-making landscape.

The examples also clarify the meaning of the bias measures introduced above. In the symmetric case both biases vanish, whereas in the presence of internal or external bias, $h_{\rm int}$ or $h_{\rm ext}$, finite $\Delta\phi$ and $\Delta F$ quantify, respectively, the imbalance in valence intensity and in stabilization between the competing affective branches. 
Such weighting may indicate a tendency toward negative or positive bias. In particular, a recurrent privileging of the unpleasant branch provides a formal representation of a bias toward negatively valence affective states. Such a bias may contribute to psychological constructs related to negative affect, including increased propensity toward unpleasant emotional experiences such as sadness, anxiety, and irritability\cite{Brandes_Chapter2018, Bernstein_BehaviorTherapy2009}. 
The bias measures $\Delta\phi$ and $\Delta F$ therefore provide a simple formal language for discussing how evaluative organization becomes stably oriented toward negative or positive affective states, and how this orientation may become pathological when it rigidifies across contexts.

\section{Sense-making landscapes and affective regimes}
\label{sec:landscapes}

Building on the qualitative discussion of affect in Sec.~\ref{sec:affect}, we now examine how the sense-making landscape evolves between the high-$T$ and low-$T$ regimes as the sensory context becomes progressively more informative about the sensation, i.e., as the contextual parameter $T$ is lowered. We identify the characteristic changes in the available affective states and their stability, showing how uncertainty, sensitization, consolidation, mixed-valence states, and history-dependent responses emerge from the evolution of the landscape, and how internal and external biases modify their symmetry and stability.
The corresponding landscape and bifurcation diagrams provide a visual representation of this evolution.

Readers who are satisfied with the high-$T$ and low-$T$ limits discussed above, and do not require the detailed landscape evolution connecting them, may proceed directly to Sec.~\ref{sec:valence_arousal}. In the following, we keep the mathematical analysis to the minimum required to identify the stationary affective states, their stability, and the characteristic contextual thresholds. The complete derivation of the stationary solutions and thresholds, together with the detailed stability analysis, is provided in Appendix~\ref{app:landscape}.

\subsection{Criteria for the landscape analysis}

Before discussing the different regimes, we introduce the basic quantities used to identify the affective states represented in the landscapes and bifurcation diagrams. For the sense-making potential of Eq.~\eqref{eq:model}, a stationary state $\bar{\phi}$ satisfies
\begin{equation}
\left.
\frac{\partial F}{\partial\phi}
\right|_{\bar{\phi}}
=
-h_{\rm ext}
+a\bar{\phi}
+h_{\rm int}\bar{\phi}^{2}
+b\bar{\phi}^{3}
=0.
\label{eq:stationary_general}
\end{equation}
These solutions define the branches followed in the bifurcation diagrams. Their local stability is determined by the curvature of the landscape,
\begin{equation}
\left.
\frac{\partial^{2}F}{\partial\phi^{2}}
\right|_{\bar{\phi}}
=
a+2h_{\rm int}\bar{\phi}
+3b\bar{\phi}^{2}.
\label{eq:curvature_general}
\end{equation}
A positive curvature identifies a locally stable minimum, whereas a negative curvature identifies an unstable stationary state. When the curvature vanishes, the local stability of a stationary branch can change.

For any locally stable minimum we define the local susceptibility to the external bias as
\begin{equation}
\chi(\bar{\phi})
=
\frac{\partial\bar{\phi}}{\partial h_{\rm ext}}.
\label{eq:susceptibility}
\end{equation}
Differentiating the stationary condition, Eq.~\eqref{eq:stationary_general}, with respect to $h_{\rm ext}$ gives
\begin{equation}
\chi(\bar{\phi})
=
\frac{1}
{a+2h_{\rm int}\bar{\phi}+3b\bar{\phi}^{2}}
=
\left[
\left.
\frac{\partial^{2}F}{\partial\phi^{2}}
\right|_{\bar{\phi}}
\right]^{-1}.
\label{eq:susceptibility_curvature}
\end{equation}
where in the last equality we have recognized the curvature defined in Eq.~\eqref{eq:curvature_general}. When $\bar{\phi}$ is the global minimum, $\chi$ coincides with the equilibrium susceptibility; for a metastable minimum, it should instead be understood as a local response. The above relation provides a direct geometrical criterion for interpreting the landscapes below. Shallow minima have a large susceptibility and are therefore easily displaced by external information, whereas strongly curved minima correspond to more robust affective states. As the curvature of a stable minimum approaches zero, the susceptibility diverges.

In the following, as $T$ is varied, we identify a set of characteristic thresholds marking qualitative changes in the landscape, such as the appearance of inflection points or new stationary states, the loss of local stability of an existing state or an exchange of global stability between competing minima. We will associate the various temperature ranges with concepts from cognitive science. In particular, the notions of sensitization and consolidation have a long tradition in psychology and cognitive science, describing how historical valuations can facilitate subsequent responses and how emerging patterns become stabilized \cite{Postman1948, Svenson1992, LewisGranic2000}. As we show below, these processes have a direct counterpart in the sense-making landscape: \textit{sensitization} describes the progressive development of an affective state, whereas \textit{consolidation} corresponds to its establishment as the globally dominant state and its subsequent stabilization as the corresponding minimum deepens. This usage should be distinguished from sensory or pain sensitization in neurophysiology and from memory consolidation in cognitive neuroscience.

\subsection{Neutrality, uncertainty, and degenerate polarized states in the absence of external and internal bias}

\begin{figure}[tbh]
\centering
\includegraphics[width=0.92\linewidth]{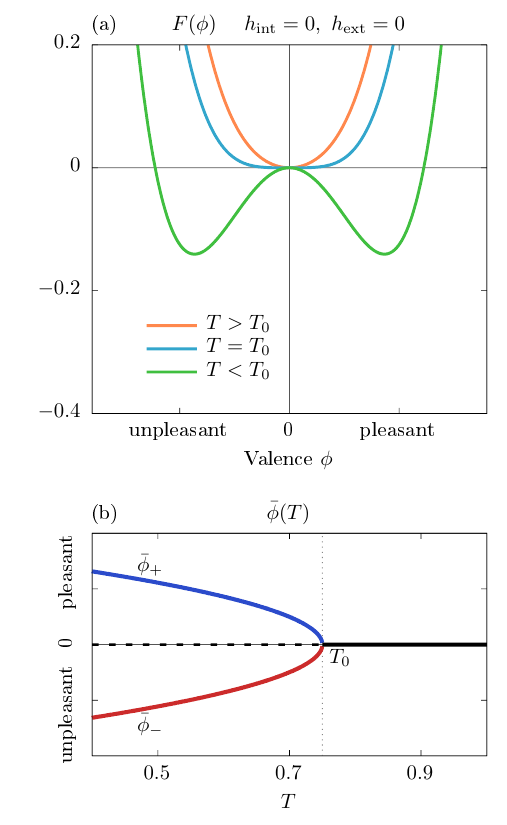}
\caption{Sense-making landscape and bifurcation structure in the symmetric case, $h_{\rm int}=h_{\rm ext}=0$.
(a) Effective potential $F(\phi)$ for $T>T_0$ (orange), $T=T_0$ (light blue), and $T<T_0$ (green). As $T$ is lowered, the landscape evolves from a neutral state above $T_0$, corresponding to lack of relevance of the sensation, to an uncertainty regime at $T_0$, where the curvature vanishes and the susceptibility diverges, and finally to two degenerate pleasant and unpleasant minima below $T_0$. Their increasing stabilization as $T$ is lowered corresponds to consolidation. 
(b) Stationary solutions $\bar\phi(T)$. Blue, red, and black denote pleasant,unpleasant, and neutral branches. Solid and dashed lines indicate stable minima and unstable maxima, respectively; thicker solid lines identify the adapted (global) minimum, and dotted vertical lines mark characteristic temperatures. For $T>T_0$, the neutral branch is stable; below $T_0$, it becomes an unstable maximum, while two degenerate polarized minima emerge continuously. The adapted realization spontaneously selects either valence.}
\label{fig:landscape_symmetric}
\end{figure}

We begin with the symmetric landscape, $h_{\rm int}=h_{\rm ext}=0$, shown in Fig.~\ref{fig:landscape_symmetric}. Panel (a) illustrates how the sense-making landscape evolves as $T$ is lowered, while panel (b) shows the corresponding  stationary solutions and their stability.

The general stationary condition in Eq.~\eqref{eq:stationary_general} in the absence of internal or external bias reduces to
\begin{equation}
\left.\frac{\partial F}{\partial\phi}\right|_{\bar{\phi}}
=
\bar{\phi}\left(a+b\bar{\phi}^2\right)=0.
\end{equation}
It always admits the neutral solution $\bar{\phi}_0=0$ and two additional polarized solutions,
\begin{equation}
\bar{\phi}_{\pm}=\pm\sqrt{-\frac{a}{b}},
\label{eq:symm_pol}
\end{equation}
for $a<0$. Their stability follows from Eq.~\eqref{eq:curvature_general}, which becomes
\begin{equation}
\left.\frac{\partial^2F}{\partial\phi^2}\right|_{\bar{\phi}}
=
a+3b\bar{\phi}^2.
\label{eq:curvature_pol}
\end{equation}

\paragraph*{\bf Neutral state}
At high temperature, $T>T_0$, we have $a=a_0(T-T_0)>0$. The neutral solution has positive curvature, $\partial^2F/\partial\phi^2=a>0$, and is the only stable minimum of the landscape, as shown by the single minimum of the orange curve in Fig.~\ref{fig:landscape_symmetric}(a). The sensation remains neutrally evaluated and, since $F(\bar{\phi}_0)=0$, it has zero arousal. Cognitively, the neutral state represents the lack of relevance of the sensation.

\paragraph*{\bf Uncertainty regime}
As $T$ is lowered toward $T_0$, the curvature of the neutral minimum progressively decreases and vanishes at $T=T_0$, where $a=0$. The landscape therefore becomes locally flat around the neutral state, as visible in the light-blue curve of Fig.~\ref{fig:landscape_symmetric}(a). The same threshold appears in Fig.~\ref{fig:landscape_symmetric}(b) as the bifurcation point at which the neutral branch loses stability as $T$ is lowered and the two polarized branches emerge.

The flattening of the landscape also signals a high sensitivity to external bias. Since the susceptibility is the inverse local curvature, along the neutral branch
\begin{equation}
\chi(\bar{\phi}_0)=1/a,
\end{equation}
and therefore diverges as $T\rightarrow T_0^{+}$. Even a very weak external bias can thus produce a large displacement of the evaluation toward one valence sector.

These properties provide a natural representation of an {\it uncertainty} regime: the valence orientation is not yet stabilized, and the evaluative organization is highly responsive both to changes in the sensory context and to socially mediated perturbations. Cognitively, this suggests that uncertainty can also provide a window of enhanced receptivity: when the affective evaluation is only weakly stabilized by the sensory context, external information can have a particularly strong impact. Such an open and flexible state may also be associated with non-ordinary states of consciousness, including certain meditative states characterized by reduced prior precision and psychedelic experiences\cite{Carhart-Harris_2023}.

\paragraph*{\bf Polarized regime and consolidation}
For $T<T_0$, $a=a_0(T-T_0)<0$, and the neutral branch becomes unstable. The two polarized solutions of Eq.~\eqref{eq:symm_pol}, instead, have positive curvature, $\partial^2F/\partial\phi^2=-2a>0$, and therefore correspond to stable degenerate minima. We identify them with the pleasant and unpleasant states,
\begin{equation}
\bar{\phi}_{pl}=\bar{\phi}_{+},
\qquad
\bar{\phi}_{unpl}=\bar{\phi}_{-}.
\end{equation}
As shown in Figure~\ref{fig:landscape_symmetric}(b) the two stable branches continuously emerge from the neutral solution at $T_0$ and separate as $T$ is lowered, with their valence intensity growing as $(T_0-T)^{1/2}$. This is the standard mean-field behavior of a continuous, second-order transition, corresponding to the pitchfork bifurcation of the symmetric Landau theory.

The green low-$T$ landscape in Fig.~\ref{fig:landscape_symmetric}(a) shows that the two minima have equal depth and hence equal arousal,
\begin{equation}
A=-F(\bar{\phi}_{pl})=-F(\bar{\phi}_{unpl})
=
\frac{a^2}{4b},
\end{equation}
which grows as $\sim (T_0-T)^2$ as $T$ is lowered further below $T_0$.
The coexistence of these two stable affective states of opposite valence therefore constitutes the symmetric, degenerate realization of the mixed affective configuration introduced in Sec.~\ref{sec:affect}.

The susceptibility of either polarized branch is $\chi=-1/(2a)$ and therefore diverges as $T\rightarrow T_0^{-}$. However, as $T$ is further reduced, the curvature of the minima increases and the susceptibility progressively decreases. Cognitively, this reduction can be interpreted as a process of {\it consolidation}: once one of the polarized evaluations is selected, it becomes increasingly robust and less easily displaced by weak socially mediated or external cues.

\subsection{{\bf Biased sensitization and consolidation: the effect of internal and external biases}}

\subsubsection{{\bf Internal bias: $h_{\rm int}\neq0$, $h_{\rm ext}=0$}}

\begin{figure}[tbh]
\centering
\includegraphics[width=0.92\linewidth]{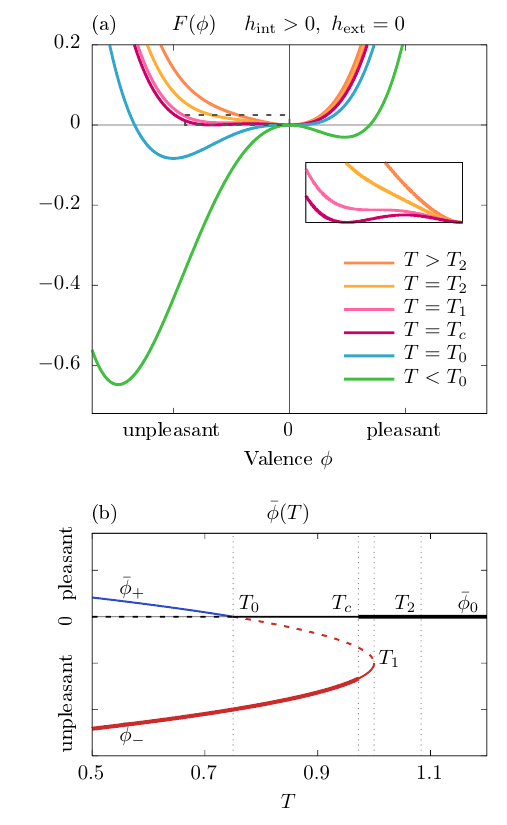}
\caption{Sense-making landscape and bifurcation structure for
$h_{\rm int}>0$ and $h_{\rm ext}=0$.
(a) $F(\phi)$ across the characteristic temperatures; colors identify the
temperatures as indicated in the legend, and the inset enlarges the boxed region
near $F=0$. Upon cooling, $T_2$ marks the first deformation toward the internally
favored unpleasant sector, while at $T_1$ a latent unpleasant minimum appears,
signaling sensitization. It remains secondary until $T_c$, where it becomes
degenerate with the neutral state and takes over as the adapted state, marking
consolidation. At $T_0$ the neutral state loses stability and a secondary pleasant
minimum emerges.
(b) Corresponding stationary branches, with the conventions of
Fig.~\ref{fig:landscape_symmetric}. The change in line thickness at $T_c$
highlights the switch of the adapted state from neutral to unpleasant.}
\label{fig:landscape_internal}
\end{figure}

The affective evolution becomes richer once a finite internal bias is introduced. Figure~\ref{fig:landscape_internal} shows how the corresponding
landscape and stationary branches evolve as $T$ is varied for
$h_{\rm int}>0$ and $h_{\rm ext}=0$; changing the sign of $h_{\rm int}$ simply
exchanges the pleasant and unpleasant sectors.

In the presence of the internal bias, the stationary condition in Eq.~\eqref{eq:stationary_general} becomes
\begin{equation}
\left.\frac{\partial F}{\partial\phi}\right|_{\bar{\phi}}
=
\bar{\phi}\left(a+h_{\rm int}\bar{\phi}+b\bar{\phi}^2\right)=0.
\label{eq:stationary_hint}
\end{equation}
The neutral solution $\bar{\phi}_0=0$ therefore remains a stationary state at any $T$, while two additional stationary branches can appear,
\begin{equation}
\bar{\phi}_{\pm}
=
\frac{-h_{\rm int}\pm
\sqrt{h_{\rm int}^2-4ab}}{2b}.
\label{eq:biased_stationary}
\end{equation}

Unlike in the symmetric case, the nonzero branches do not emerge directly at $T_0$.
As made explicit by the landscape sequence in
Fig.~\ref{fig:landscape_internal}(a) and by the bifurcation structure in
panel (b),  the internally biased landscape instead evolves through a sequence of characteristic temperatures,
\begin{equation}
T_2>T_1>T_c>T_0.
\end{equation}
The three additional thresholds, whose derivation is given in Appendix~\ref{app:landscape}, can be written as
\begin{equation}
T_i
=
T_0+c_i\frac{h_{\rm int}^2}{a_0b},
\qquad
c_2=\frac{1}{3},\quad
c_1=\frac{1}{4},\quad
c_c=\frac{2}{9}.
\label{eq:characteristic_scales}
\end{equation}
The characteristic temperatures are shown in Fig.~\ref{fig:landscape_internal}(a), with colors identifying the corresponding landscapes and the inset highlighting the shallow structure near $F=0$. $T_2$ (yellow) marks the first deformation of the landscape toward the sector favored by the internal bias, through the development of an inflection point; $T_1$ (pink) marks the appearance of the two nonzero stationary branches $\bar{\phi}_{\pm}$; $T_c$ (magenta) is the coexistence threshold at which the neutral minimum and the unpleasant minimum have equal depth and exchange global stability; and $T_0$ (light blue) remains the threshold at which the neutral stationary state $\bar{\phi}_0$ loses local stability.
Panel (b) shows the corresponding reorganization of the stationary branches and their stability.

As can be seen from Eq.~\eqref{eq:characteristic_scales}, the three additional thresholds are shifted above $T_0$ on a scale set by $h_{\rm int}^2/(a_0b)$. Their displacement therefore depends on the magnitude of the internal bias but not on its sign, while it is also modulated by the Landau coefficients $a_0$ and $b$, which encode properties inherited from the microscopic network of interacting vital norms~\cite{Fanfarillo_RSOS2026}. Cognitively, this means that the same strength of learned evaluation can reorganize the affective landscape under different sensory conditions in different individuals, depending on the parameters $a_0$ and $b$.

\paragraph*{\bf Onset of bias and sensitization}
For $T>T_2$, the landscape contains only the neutral minimum $\bar{\phi}_0=0$, which corresponds to vanishing valence and arousal. At $T=T_2$, an inflection point develops in the unpleasant sector favored by $h_{\rm int}$. No additional stationary state is present yet, but the landscape has begun to deform asymmetrically. This initial deformation, which is subtle on the scale of the full
landscape, is made visible by the inset of Fig.~\ref{fig:landscape_internal}(a).
For $T_1<T<T_2$, this deformation becomes progressively more pronounced while $\bar{\phi}_0$ remains the only minimum.
At $T=T_1$, the two nonzero stationary branches of Eq.~\eqref{eq:biased_stationary} appear with vanishing local curvature and therefore divergent susceptibility.

As seen directly in Fig.~\ref{fig:landscape_internal}(b), below $T_1$ the branch $\bar{\phi}_{-}$ becomes a local unpleasant minimum, $\bar{\phi}_{unpl}=\bar{\phi}_{-}$, whereas $\bar{\phi}_{+}$ is an unstable maximum separating it from the neutral minimum. Throughout the interval $T_c<T<T_1$, the neutral state remains globally adapted, while $\bar{\phi}_{unpl}$ represents a locally stable but secondary affective possibility.

We interpret this regime as a stage of {\it sensitization}: an unpleasant evaluation is already encoded in the landscape as a latent state but has not yet become the dominant adapted state.
Since the curvature of the emerging branch vanishes at $T_1$, its susceptibility diverges there, indicating an extreme sensitivity to weak external perturbations at the onset of sensitization.

\paragraph*{\bf Coexistence threshold and consolidation}
At $T=T_c$, the unpleasant and neutral minima have equal depth, $F(\bar{\phi}_{unpl})=F(\bar{\phi}_0)=0$, and therefore equal, vanishing arousal,
as shown most clearly in the enlarged region of
Fig.~\ref{fig:landscape_internal}(a).
Unlike $T_1$, this threshold is not associated with vanishing curvature or divergent susceptibility; instead, it marks an exchange of global stability between the two competing minima. At coexistence, the polarized minimum is already located at the finite value $\bar{\phi}_{unpl}=-2h_{\rm int}/(3b)$. Thus, as $T$ is lowered through $T_c$, the globally adapted state changes discontinuously from the neutral solution $\bar{\phi}=0$ to a finite unpleasant valence.
In the bifurcation diagram this exchange is highlighted by the
switch in line thickness between the neutral and unpleasant branches at $T_c$. The transition is therefore first order, in contrast to the continuous transition of the symmetric case.

From the affective perspective, the latent unpleasant evaluation present above $T_c$ now becomes the dominant adapted state, while the neutral state remains locally stable down to $T_0$. We interpret this evolution as {\it consolidation}: as $T$ is lowered further toward $T_0$, the unpleasant minimum deepens, and the corresponding affective organization becomes progressively more stabilized.

\paragraph*{\bf Biased mixed-valence regime}
At $T=T_0$, the branch $\bar{\phi}_{+}$ reaches the neutral solution and its local curvature vanishes. The neutral branch therefore loses local stability, with a corresponding divergence of its susceptibility.
For $T<T_0$, $\bar{\phi}_{+}$ moves into the positive-valence sector and becomes the local pleasant minimum, $\bar{\phi}_{\rm pl}=\bar{\phi}_{+}$, while $\bar{\phi}_{\rm unpl}=\bar{\phi}_{-}$ remains the dominant unpleasant state. The landscape therefore supports two locally stable affective states of opposite valence. This realizes an asymmetric mixed-affect configuration: a dominant unpleasant state coexists with a secondary pleasant one. In contrast to the symmetric case, both the valence bias $\Delta\phi$ and the potential bias $\Delta F$ are finite, quantifying the imbalance in valence intensity and stabilization between the competing affective states.

\paragraph*{\bf Metastability and history dependence}
The coexistence of locally stable minima also introduces the possibility of metastability and history-dependent behavior. In the interval $T_0<T<T_1$, both the neutral and unpleasant states are locally stable, while their global stability is exchanged at $T_c$.
This is represented in Fig.~\ref{fig:landscape_internal}(b) by
the coexistence of two solid branches, with the thicker segment identifying the
globally adapted state on either side of $T_c$. The state occupied by the system can therefore depend not only on the current sensory context but also on the trajectory through which that context was reached.
For example, upon increasing $T$ from the consolidated unpleasant regime, the unpleasant evaluation can persist metastably above $T_c$, even after the neutral state has become globally favored, and disappear only at $T_1$ when its local minimum loses stability. Cognitively, this provides a possible mechanism for persistence or inertia of an established affective evaluation. A direct clinical interpretation, however, would require an explicit dynamical description and empirical validation.

\subsubsection{{\bf External bias: $h_{\rm ext}\neq0$, $h_{\rm int}=0$}}

\begin{figure}[tbh]
\centering
\includegraphics[width=0.92\linewidth]{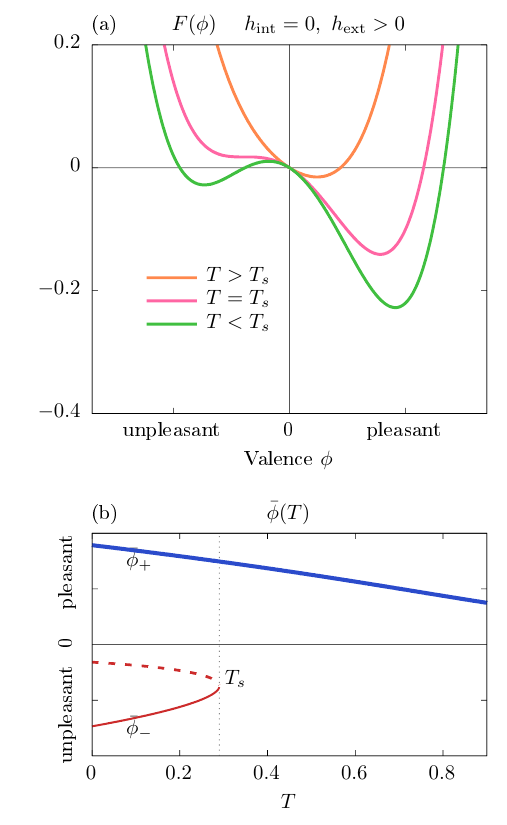}
\caption{Sense-making landscape and bifurcation structure for
$h_{\rm int}=0$ and $h_{\rm ext}>0$.
(a) $F(\phi)$ above, at, and below $T_s$. The external bias selects a pleasant adapted state already at high $T$ and continuously guides it away from neutrality as $T$ is lowered, i.e., as the sensory context becomes more informative about the sensation. At $T_s$ an unpleasant minimum--maximum pair appears, introducing a latent opposite-valence state.
(b) Corresponding stationary branches, with the conventions of Fig.~\ref{fig:landscape_symmetric}. The pleasant branch remains the adapted global minimum throughout, whereas the unpleasant minimum below $T_s$ is a locally stable latent state. Thus $T_s$ changes the available affective states without producing a transition of the adapted state.}
\label{fig:landscape_external}
\end{figure}

We now consider the complementary case of a finite external bias, $h_{\rm ext}\neq0$, with $h_{\rm int}=0$. For definiteness, we take $h_{\rm ext}>0$, as shown in Fig.~\ref{fig:landscape_external}; changing its sign simply exchanges the pleasant and unpleasant sectors. The stationary condition in Eq.~\eqref{eq:stationary_general} becomes
\begin{equation}
\left.\frac{\partial F}{\partial\phi}\right|_{\bar{\phi}}
=
-h_{\rm ext}+a\bar{\phi}+b\bar{\phi}^3=0.
\label{eq:stationary_hext}
\end{equation}
The main difference from the internal-bias case is immediately apparent: $\bar{\phi}=0$ is no longer a stationary solution. As already discussed in Sec.~\ref{sec:affect}, the external bias selects one valence sector already at high $T$, while at sufficiently low $T$ two inequivalent minima of opposite valence can coexist. Figure~\ref{fig:landscape_external}(a) shows the deformation of the landscape upon lowering $T$, while panel (b) follows the corresponding stationary states and their stability.

\paragraph*{\bf Externally guided affective state}
At high $T$, the landscape contains a single minimum displaced toward the valence sector favored by $h_{\rm ext}$, corresponding here to a pleasant adapted state $\bar{\phi}_{pl}$. This displacement from neutrality is already visible in the high-$T$ orange landscape of Fig.~\ref{fig:landscape_external}(a). Its valence and arousal are finite, although both tend to zero in the limit of very large $T$. As $T$ is lowered, the minimum moves progressively farther from neutrality and becomes deeper, so that both valence intensity and arousal increase continuously.

This evolution remains smooth across $T_0$. As visible from the landscape, $\bar{\phi}_{pl}$ retains finite valence, finite arousal, and positive curvature at $T=T_0$. Thus, unlike in the symmetric case, $T_0$ no longer marks a loss of local stability or a divergent susceptibility of the adapted state. More generally, at fixed $h_{\rm ext}\neq0$ the globally adapted state evolves continuously with $T$, so that no phase transition occurs as the sensory context is varied. Below $T_0$ the shape of the landscape begins to change through the development of two inflection points, while $\bar{\phi}_{pl}$ remains the only stable minimum.

\paragraph*{\bf Emergence of a latent opposite-valence state}
Upon further lowering $T$, Fig.~\ref{fig:landscape_external}(a) shows a second qualitative change at a threshold $T=T_s$, (pink), where two additional stationary points appear in the opposite-valence sector. This does not correspond to a phase transition, since the globally adapted pleasant state evolves continuously through $T_s$. At the threshold, the emerging stationary points have vanishing curvature and therefore divergent susceptibility. Their explicit stability analysis and the expression for $T_s$ are given in Appendix~\ref{app:landscape}.
For $T<T_s$, they separate into a local unpleasant minimum, $\bar{\phi}_{unpl}$, and an unstable maximum, while $\bar{\phi}_{pl}$ remains the globally stable minimum. The corresponding minimum--maximum pair is visible in the bifurcation diagram of Fig.~\ref{fig:landscape_external}(b), while the pleasant branch remains thick throughout, emphasizing that it remains the adapted state. The unpleasant state therefore represents a latent state. Immediately after its appearance it lies above $F=0$ and does not yet carry positive arousal; upon further lowering $T$, it crosses below $F=0$ and acquires finite arousal while remaining secondary to $\bar{\phi}_{pl}$.

This evolution explains how the mixed-valence configuration develops in the presence of an external bias. In contrast to the sensitization--consolidation sequence generated by $h_{\rm int}$, the externally favored valence sector remains globally stable throughout the evolution, while the opposite-valence state appears only at lower $T$ as a latent alternative. At sufficiently low $T$, the landscape therefore contains a dominant pleasant state and a secondary unpleasant state with different valence intensities and arousal.

\paragraph*{\bf Metastability and history dependence}
At fixed $h_{\rm ext}\neq0$, the globally adapted state evolves continuously with $T$, although a system occupying the latent minimum may remain metastably trapped there until that minimum disappears. A different behavior emerges when $h_{\rm ext}$ itself is varied. For fixed $T<T_0$, changing the sign of the external bias exchanges the global stability of the two opposite-valence minima. At $h_{\rm ext}=0$ the two minima are degenerate, and the globally adapted state changes discontinuously between finite pleasant and unpleasant valence. The transition driven by $h_{\rm ext}$ is therefore first order.

The coexistence of the two locally stable states also allows for history-dependent behavior. An affective state induced by external information may remain locally stable when that information is weakened and even after its sign is reversed, disappearing only when the corresponding minimum loses stability. The resulting evaluation can therefore depend not only on the current external cue, but also on the sequence of external information through which the present state was reached. Cognitively, this provides a possible mechanism by which socially or clinically communicated information may leave a persistent affective trace, potentially relevant to placebo and nocebo effects, where externally provided expectations can modify the evaluation of bodily sensations and persist beyond the immediate context in which they were induced. 

\subsubsection{{\bf Combined internal and external bias}}

\begin{figure}[tbh]
\centering
\includegraphics[width=0.92\linewidth]{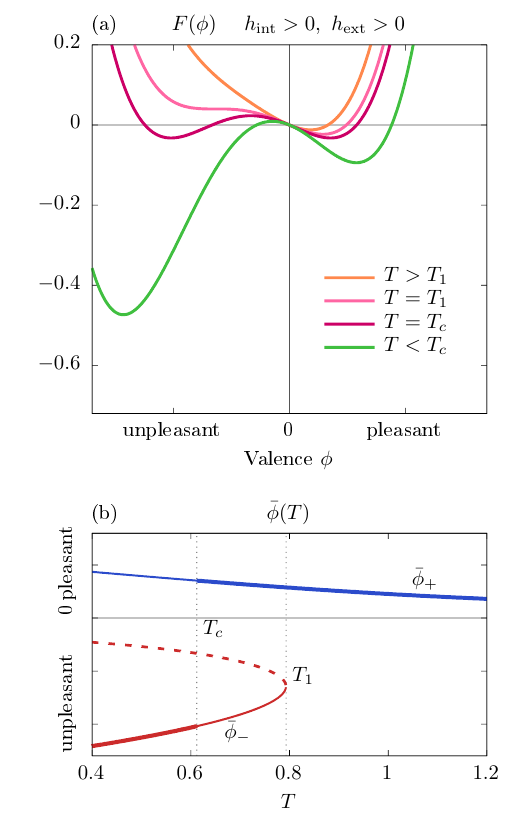}
\caption{Sense-making landscape and bifurcation structure for competing internal and external biases, $h_{\rm int}>0$ and $h_{\rm ext}>0$.
(a) $F(\phi)$ across $T_1$ and $T_c$. Here the external bias favors pleasant valence, whereas the internal bias favors unpleasant valence. The adapted state is initially pleasant; at $T_1$ a latent unpleasant minimum appears, and at $T_c$ the two minima become degenerate and exchange global stability. Below $T_c$ the internally favored unpleasant evaluation becomes adapted and progressively consolidates.
(b) Corresponding stationary branches, with the conventions of
Fig.~\ref{fig:landscape_symmetric}. The thickness switch at $T_c$ makes the change of adapted state explicit, while the competing stable branch remains metastable.}
\label{fig:landscape_competing}
\end{figure}

The qualitative interplay between $h_{\rm int}$ and $h_{\rm ext}$ was introduced in Sec.~\ref{sec:affect}.
Figure~\ref{fig:landscape_competing} illustrates a representative competing case, $h_{\rm int}>0$ and $h_{\rm ext}>0$, in which the internal bias favors unpleasant valence while the external bias favors pleasant valence. We now further explore the consequences of their combined action.

\paragraph*{\bf Modified contextual thresholds}
The characteristic thresholds retain the same qualitative meaning introduced for $h_{\rm ext}=0$. In Fig.~\ref{fig:landscape_competing}(a), we explicitly show $T_1$ (pink) that marks the appearance of the additional unpleasant minimum and unstable maximum, and $T_c$ (magenta) the coexistence threshold at which the two minima have equal depth and exchange global stability (as emphasized also in panel (b)).

A finite external bias affects the characteristic thresholds selectively: $T_2$ remains unchanged, since $h_{\rm ext}$ does not enter the curvature of the landscape, Eq.~\eqref{eq:curvature_general}, whereas $T_1$ and $T_c$ are shifted according to
\begin{align}
T_i(h_{\rm int},h_{\rm ext})
\sim
T_0
+c_i\frac{h_{\rm int}^2}{a_0b}
- d_i \frac{b h_{\rm ext}}{a_0 h_{\rm int}},
\label{eq:Tc_combined}
\end{align}
with $d_1=2$, $d_c=3$, and the prefactors $c_i$ the same as in
Eq.~\eqref{eq:characteristic_scales} (full derivation in
Appendix~\ref{app:combined}). As in the purely externally biased case, $T_0$ no longer corresponds to the loss of stability of a neutral state, since $\bar{\phi}=0$ is not stationary for $h_{\rm ext}\neq0$.

Equation~\eqref{eq:Tc_combined} shows that the shift of $T_1$ and $T_c$ is controlled by the strength of the external bias relative to the internal one, through the ratio $h_{\rm ext}/h_{\rm int}$, and changes sign depending on whether
the two biases reinforce or oppose each other. In particular, the shift of $T_c$ provides a transparent measure of how external information modifies the competition between affective states already shaped by the internal bias.
For $h_{\rm int}>0$, which favors the unpleasant sector, a positive $h_{\rm ext}$ opposes this tendency and lowers $T_c$, so that a more informative sensory context is required for the internally favored state to become dominant.
A negative $h_{\rm ext}$ instead reinforces the unpleasant tendency and shifts $T_c$ upward, allowing $\bar{\phi}_{unpl}$ to become dominant already in a less
informative context.

\paragraph*{\bf Affect at coexistence}
The displacement of the competing minima also changes the affective organization at coexistence. Depending on the relative direction and strength of the two biases, the two coexisting states may have opposite valence or may lie within the same valence sector with different valence intensities.
In the example of Fig.~\ref{fig:landscape_competing}(a), the magenta landscape at $T_c$ shows the two opposite-valence minima at equal depth, while panel (b) highlights their exchange of global stability through the switch in line thickness at $T_c$. Since the adapted state changes discontinuously between two finite-valence minima, the transition at $T_c$ is first order.

The arousal at coexistence also becomes finite. In contrast to the internally biased case, where both competing states have zero arousal at $T_c$, for $h_{\rm ext}\neq0$ the two minima coexist at the same finite arousal. This coexistence arousal scales as $(h_{\rm ext}/h_{\rm int})^2$, with the complete expression given in Appendix~\ref{app:landscape}, Eq.~(\ref{eq:app_Ac}).

Cognitively, this means that the contextual conditions for a change of the dominant affective state, as well as the arousal associated with that competition, depend on the strength of current external information relative to the pre-existing learned tendency, and on whether the two reinforce or oppose each other. 

\paragraph*{\bf Metastability, and history dependence}
At sufficiently low $T$, locally stable states can occupy opposite valence sectors, producing a biased mixed-affect configuration: a dominant unpleasant state coexists with a secondary pleasant one. This regime is visible in Fig.~\ref{fig:landscape_competing}: the low-$T$ green landscape contains two minima of opposite valence, while panel (b) shows the unpleasant branch as the adapted state and the pleasant branch as the secondary stable state. Their valence intensity, relative stabilization, and arousal are jointly shaped by $h_{\rm int}$ and $h_{\rm ext}$.

The resulting multistable landscape also allows for metastability and hysteretic responses when the external bias is varied. Because $h_{\rm int}$ already favors one sector of the landscape, this hysteresis is generally asymmetric: the external bias required to destabilize an established state need not be the same in the two directions. Cognitively, the same external information can therefore have different effects depending on the learned valuation on which it acts, making an established affective tendency easier to reinforce than to reverse and providing a possible mechanism for resistance to opposing external information. A closely related asymmetry has been documented for pain expectations, which become self-reinforcing because expectation-congruent experiences are weighted more strongly during learning than incongruent ones\cite{jepma-2018}. This individual dependence may also be relevant to personalized medicine, as the same external intervention, represented in the model through the parameter $h_{ext}$, may produce different responses depending on the individual's internal bias $h_{\rm int}$, and the intrinsic parameters $a_0$, $b$, and $T_0$, shaping their affective landscape.

\section{Emergent valence--arousal relations}
\label{sec:valence_arousal}

Having characterized how the minima of the sense-making landscape emerge, evolve, and change their stability across the different affective regimes, we now examine the corresponding organization in the valence--arousal plane. The sense-making framework does not prescribe a valence--arousal relation as an independent phenomenological input. Rather, each locally stable minimum defines an affective state through its position $\bar{\phi}$ and its depth $A(\bar{\phi})$. As the parameters controlling the landscape are varied, the minima change their position and depth, and the corresponding affective states trace trajectories in the valence--arousal plane. The resulting relation is therefore not expected to be universal: its form depends on which parameters vary, which remain approximately fixed, and which stationary branches are sampled.

As discussed in Sec.~\ref{sec:model}, $T$, $h_{\mathrm{int}}$, and $h_{\mathrm{ext}}$ represent distinct sources of contextual and evaluative variation. Changes in these parameters consequently deform the landscape in different ways and generate different valence--arousal relations. Varying $T$ while keeping $h_{\mathrm{int}}$ and $h_{\mathrm{ext}}$ fixed isolates how changes in contextual noise modify valence and arousal. By contrast, varying $h_{\mathrm{int}}$ or $h_{\mathrm{ext}}$ probes how learned history or external influence reshapes the valence--arousal relation.

\begin{figure*}
\centering
\includegraphics[width=0.99\textwidth]{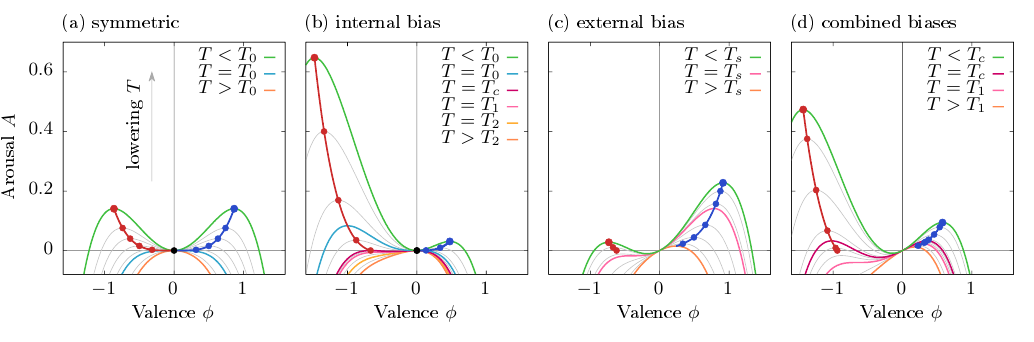}
\caption{Valence--arousal relations generated by the sense-making landscape for representative cases: (a) symmetric, (b) internal bias, (c) external bias, and (d) competing internal and external biases. Arousal is the depth of a stable minimum relative to $F=0$, and is plotted against valence $\phi$. Colored curves correspond to representative temperatures, using the same temperature color code adopted in Sec.~\ref{sec:landscapes}, while light-gray curves show intermediate temperatures. Red, blue, and black dots mark the stable unpleasant, pleasant, and neutral affective states associated with the corresponding minima of the landscape. The connecting lines trace the evolution of these states as $T$ is lowered, i.e. as the sensory context becomes more informative.
All cases present an upward-open organization of the valence--arousal relation under variation of $T$: as stable states move away from neutrality, their arousal generally increases. In the symmetric case, panel~(a), the relation is symmetric: the pleasant and unpleasant branches emerge continuously below $T_0$ and remain degenerate. In the presence of bias, panels~(b)--(d), this symmetry is broken, with the direction of the asymmetry determined by the sign of the bias. For the cases shown here, $h_{\rm int}>0$ in panel~(b) favors the unpleasant sector: a latent unpleasant minimum first develops in the landscape and enters the valence--arousal representation, subsequently becoming the dominant affective branch, while a secondary pleasant branch emerges at lower $T$. For $h_{\rm ext}>0$, panel~(c), the pleasant branch is dominant already at high $T$ and evolves continuously, whereas an unpleasant minimum appears only at lower $T$. In panel~(d), the two biases compete: the externally favored pleasant branch is initially dominant, while an internally favored unpleasant minimum develops as the context becomes more informative, enters the valence--arousal representation as a secondary affective state, and eventually becomes dominant.}
\label{fig:Vshape}
\end{figure*}

\subsection*{The upper-open relation generated by contextual variation}

To illustrate the general construction, we consider a theoretical trajectory obtained by varying $T$ while keeping $h_{\mathrm{int}}$, $h_{\mathrm{ext}}$, and the structural coefficients of the sense-making potential fixed. This provides the most direct connection with the previous section, where the evolution of the sense-making landscapes with $T$ was analyzed in detail. As $T$ changes, the stationary minima move in valence and change their depth. Those crossing the neutral reference $F=0$ define affective states and trace trajectories in the valence--arousal plane.
Using the stationarity condition introduced above, the arousal associated with any nonzero minimum can be written directly as
\begin{equation}
A(\bar{\phi})
=
\frac{h_{\mathrm{ext}}}{2}\bar{\phi}
+
\frac{h_{\mathrm{int}}}{6}\bar{\phi}^{3}
+
\frac{b}{4}\bar{\phi}^{4}.
\label{eq:emergent_VA}
\end{equation}
(see Appendix~\ref{app:general} for full derivation). The temperature does not appear explicitly in this expression because it determines the position of $\bar{\phi}(T)$. The internal and external biases instead control the asymmetry between the pleasant and unpleasant branches.
In Fig.~\ref{fig:Vshape}, we show the full $-F(\phi)$ curves to make their connection with the corresponding sense-making landscapes explicit, whereas the dots and connecting lines identify only affective states with zero or positive arousal.

For $h_{\mathrm{int}}=h_{\mathrm{ext}}=0$, the relation reduces to
\begin{equation}
A(\bar{\phi})=\frac{b}{4}\bar{\phi}^{4},
\end{equation}
which is symmetric under $\bar{\phi}\rightarrow-\bar{\phi}$ and produces the smooth,  upper-open shape shown in Fig.~\ref{fig:Vshape}(a). The arousal vanishes at neutral valence, $\bar{\phi}_0$, and increases equally toward pleasant and unpleasant valence $\bar{\phi}_{\pm}$. Within the quartic Landau approximation, the curve is smooth near the origin and may therefore appear U-like rather than as a geometrically sharp V. Hereafter we use the term V-shaped in the broader empirical sense of an upper-open relation in which arousal increases with valence intensity.

When $h_{\mathrm{int}}\neq0$ and $h_{\mathrm{ext}}=0$, the cubic contribution in Eq.~\eqref{eq:emergent_VA} breaks the symmetry between the two valence sectors. For the sign shown in Fig.~\ref{fig:Vshape}(b), $h_{\mathrm{int}}>0$, the unpleasant sector is preferentially stabilized. As $T$ is lowered, i.e. as the sensory context becomes more informative, an unpleasant minimum first develops in the landscape during the sensitization regime, where the neutral state is still the adapted one. Upon further lowering $T$, the unpleasant state crosses zero arousal, enters the valence--arousal representation, and progressively consolidates as the dominant affective state. A secondary pleasant affective branch emerges only below $T_0$. This results in an asymmetric V-shape of the valence--arousal relation.

A finite $h_{\mathrm{ext}}$ introduces an additional linear contribution to Eq.~\eqref{eq:emergent_VA}. For $h_{\mathrm{int}}=0$ and $h_{\mathrm{ext}}>0$, shown in Fig.~\ref{fig:Vshape}(c), the pleasant branch is selected already at high $T$ and grows continuously in valence and arousal as $T$ is lowered. An unpleasant secondary affective branch emerges only at lower $T$. Thus, socially mediated or externally supplied information can generate an asymmetric valence--arousal relation even in the absence of a learned historical bias.

When both biases are present, the valence--arousal relation reflects their interplay. In the competing case shown in Fig.~\ref{fig:Vshape}(d), current external information favors a pleasant evaluation, whereas the learned internal tendency favors an unpleasant one. As the sensory context becomes more informative, an unpleasant minimum develops in the landscape and, once sufficiently deep, enters the valence--arousal representation as a secondary affective branch that can eventually become dominant, producing a change in the prevailing affective state.

All four panels of Fig.~\ref{fig:Vshape} isolate the effect of contextual variation: $T$ is varied while $h_{\mathrm{int}}$, $h_{\mathrm{ext}}$, and the structural coefficients are kept fixed within each panel. More generally, if $T$, $h_{\mathrm{int}}$, and $h_{\mathrm{ext}}$ vary simultaneously the observed affective states need not lie on a single fixed curve, but may instead sample a broader family of landscape-generated relations.

The key result of our analysis is that the valence--arousal relation is not an independent phenomenological input. It is generated by the same stationary minima of the sense-making landscape whose temperature evolution was analyzed in Sec.~\ref{sec:landscapes}. The upper-open V-like relation is not imposed on the model but emerges as a particular projection of the sense-making landscape when contextual informativeness is varied at fixed bias. Its symmetry, asymmetry, and smoothness are controlled by the coefficients shaping the landscape. The broader prediction of the framework is that different empirical valence--arousal relations should be interpretable as projections of different trajectories through the same underlying parameter space.

\subsection*{Relation to empirical observations and experimental testing}

Classical psychophysiological studies using affective pictures often reported an upper-open V-shape, or {\it boomerang} organization of the valence--arousal plane: highly pleasant and highly unpleasant stimuli tend to elicit stronger subjective, autonomic, and behavioral responses than neutral stimuli
\cite{Lang1993LookingPictures, Bradley2001Emotion}.
However, the empirical relation is not universal. Its shape may be asymmetric and can vary across individuals, experimental protocols, stimulus sets, and cultural contexts
\cite{Ito2005Variations, Kuppens:2013aa, Kuppens_2017, Yik:2023aa}.
In particular, relations obtained from standardized affective stimuli may differ from those reconstructed from affective experiences sampled in everyday life. Recent cross-cultural evidence also indicates substantial variability, with pooled data showing an increase of arousal with positive valence that is not accompanied by an equivalent dependence on the negative side
\cite{Yik:2023aa}.

The relation generated by the theoretical $T$-scan, shown in Fig.~\ref{fig:Vshape}, reproduces the general upper-open organization observed in these studies and naturally allows for both symmetric and asymmetric branches. However, this correspondence does not imply that existing experimental protocols realize a pure variation of $T$. In classical stimulus paradigms, the sensory content of the stimulus, its contextual relevance, its learned meaning, and its motivational and cultural significance generally vary simultaneously. The observed valence--arousal relation may therefore represent a trajectory involving several parameters of the sense-making landscape rather than a change in contextual informativeness alone. In addition, the distribution of the sampled stimuli or experiences may influence the apparent shape of the relation. A dataset enriched in highly arousing pleasant and unpleasant stimuli may emphasize an upper-open pattern, whereas naturalistic sampling may produce weaker or more heterogeneous relations.

From an empirical standpoint, the framework suggests that valence and arousal should be operationalized through complementary measurements rather than identified with a single experimental readout. Subjective valence and experienced arousal are commonly assessed using self-report instruments such as the Self-Assessment Manikin and through standardized sets of affective stimulus such as the International Affective Picture System, which provide normative ratings of pleasantness, arousal, and dominance
\cite{Bradley_1994, Lang2008IAPS}.
Although the eliciting input is often an external stimulus, these ratings characterize the affective response of the observer rather than a physical property of the stimulus itself.

Self-reports may be complemented by appraisals and approach--withdrawal behavior, as well as by peripheral physiological measures including electrodermal activity, heart rate, heart-rate variability, respiration, and pupil diameter. Electrophysiological and neuroimaging measurements provide additional information about autonomic, wakeful, and affective mobilization
\cite{Bradley_2008, Beissner_2013, Satpute19Barrett}.
Within the present framework, subjective ratings, appraisals, and behavior may help constrain the selected evaluative position $\bar{\phi}$, whereas physiological and neural measurements may provide complementary empirical correlates of the landscape-defined arousal $A(\bar{\phi})$. However, these different measurements should not be assumed to be mutually equivalent or to provide a direct measurement of either model quantity.

The model also suggests parameter-specific experimental manipulations. The informativeness of the sensory context could be varied to probe the role of $T$, while prior learning and socially supplied evaluative information could be manipulated to investigate $h_{\mathrm{int}}$ and $h_{\mathrm{ext}}$, respectively. The resulting changes could be examined not only in the selected valence and arousal, but also in the asymmetry between pleasant and unpleasant states, the stability and curvature of the corresponding minima, and the susceptibility to external information.

A selective test of the valence--arousal relation generated by varying $T$ would require keeping the evaluated sensation and the relevant learned and social biases approximately fixed while systematically changing how informative the surrounding context is with respect to that sensation. Experience-sampling protocols could provide a complementary naturalistic test, provided that valence--arousal reports are accompanied by information about the sensation or event being evaluated, its contextual relevance, prior associations, and current social influences. Existing empirical evidence is therefore compatible with the landscape-based construction, but it does not yet establish that observed V-shaped relations originate from variations of $T$ alone.

\section{Discussion}
\label{sec:discussion}

\paragraph*{\bf Allostatic and enactive perspectives}

Formulating affect within the sense-making phase transition framework~\cite{Valenzuela_FrontPain2024,Fanfarillo_RSOS2026} allows us to examine more closely how this construction relates to existing accounts of affect. Affect emerges directly from the same sense-making landscape that organizes the evaluation of the sensation. 
In this respect, the construction gives formal expression to the idea of affect as an evaluative structure that guides the organism's relation to its world, in a way that is conceptually close both to Barrett's account of affect and to enactive approaches emphasizing sense-making as an organism's situated evaluation of its world~\cite{Barrett:2009aa,LFBarrettBook,Varela:2016aa,Di-Paolo:2017aa,Garcia:2025aa}.

This connection can be made more explicit by noting that Barrett's allostatic account and enactive approaches converge on the idea that affect is an organism-centered evaluative orientation. For Barrett, affect reflects the ongoing regulation of bodily resources by the organism in relation to the demands of the world; for enactivism, sense-making is the process through which an autonomous organism discloses what matters according to its norms, needs, and possibilities for action. The sense-making landscape provides a formal representation of this shared intuition by defining the space of possible evaluative orientations toward a with each minimum identifying an affective state: its position specifies valence, its depth quantifies arousal, and the global minimum identifies the adapted state.
The framework therefore connects the regulatory and relational aspects of affect within the same mathematical description, without requiring a commitment to a specific biological or cognitive implementation.

\paragraph*{\bf Affective organization and valence--arousal relation} 

Several conceptual consequences follow from this construction. First, the framework clarifies the relation between valence intensity and arousal. The two are distinct quantities: the former is measured by $|\bar{\phi}|$, while the latter is determined by the depth of the corresponding minimum. They are nevertheless not independent, since both emerge from the same sense-making landscape. This distinction is important in relation to the literature, where valence intensity and arousal have often been conflated despite evidence that they constitute distinct aspects of affective experience~\cite{Kuppens:2013aa}.

Second, the framework naturally accommodates mixed-valence states without introducing positive and negative affect as two independent dimensions. A single valence dimension can contain simultaneously a locally stable pleasant state and a locally stable unpleasant state. 
When their depths differ, the two states can be distinguished as dominant and secondary affective states.
Mixed affect therefore emerges from the availability of different states of the sense-making landscape itself. This differs from descriptions in which positive and negative affect are represented as separate dimensions~\cite{Watson1988,Russell2017}, as implemented, for example, in the Affective Ising Model~\cite{Loossens:2020aa}.

Finally, because valence and arousal are generated by the same landscape, their mutual relation also follows from its structure. The upper-open valence--arousal organization discussed in the affect literature is therefore not imposed independently on the model. As the sensory context changes, the affective states move in the valence--arousal plane, generating the relation analyzed in Sec.~\ref{sec:valence_arousal}.
More generally, the framework suggests that different empirical valence--arousal relations may reflect different ways in which contextual information, learned valuation, and external influence shape the underlying affective landscape.

\paragraph*{\bf Affective regimes and their cognitive interpretation}

A particularly useful feature of the landscape description is that it provides a natural geometric organization of different affective regimes. When the sensory context carries little information relevant to the sensation, the evaluation can remain neutral, with no stabilized pleasant or unpleasant affective orientation. As the context becomes more informative, the landscape may enter an uncertainty regime in which the evaluation is still weakly stabilized and therefore particularly sensitive to external information. In this sense, uncertainty is not simply the absence of an affective state, but a condition in which alternative evaluations can be more easily shaped by the context or by external information.

In the presence of a learned bias, the landscape distinguishes different stages in the establishment of an affective evaluation. A latent affective possibility may first appear as a secondary state without yet determining the adapted evaluation; we associate this onset with sensitization. As the sensory context becomes more informative, the latent state can become dominant. With a further increase in contextual information, the corresponding minimum becomes progressively deeper, providing a natural landscape representation of consolidation. Importantly, the contextual conditions under which these changes occur are not determined by $h_{\rm int}$ alone, but are modulated by the individual parameters $a_0$, $b$, and $T_0$ that shape the landscape. The same strength of learned valuation may become affectively relevant under different sensory conditions in different individuals. Finally, below $T_0$, a secondary minimum of opposite valence is also stabilized, leading to a biased mixed-valence regime.

External information acts differently: a finite $h_{\rm ext}$ can directly shift the adapted evaluation even when the sensory context is weak, while at sufficiently informative contexts the landscape may also support a secondary state of opposite valence.

In the presence of an internal bias $h_{\rm int}$, an external bias $h_{\rm ext}$, or both, the coexistence of alternative affective states can give rise to history dependence and hysteresis loops. An established evaluation may persist even when the current context begins to favor another possibility, so that the affective response depends not only on the present information but also on the trajectory through which the current state was reached. The learned valuation can further make this persistence asymmetric, so that the same external information may reinforce an established tendency more easily than reverse it.
This provides a possible formal language for discussing affective flexibility and persistence, as well as placebo and nocebo effects, and may also be relevant to individual differences in response to external interventions, including in the context of personalized medicine. However, a direct interpretation in specific clinical conditions requires empirical validation and an explicit dynamical description.

\paragraph*{\bf Beyond affect of an internal sensation} 

Although the present work has been formulated in terms of the evaluation of an inner sensation, the model is not restricted to interoception. Valence is constructed through the evaluation of a sensation and is not assumed to be an intrinsic property of the sensory signal itself. The formalism is therefore agnostic with respect to the sensory channel through which the relevant information is conveyed: what matters is the contextual evaluation of the sensory signal rather than its sensory origin. The same framework can therefore be extended to affect elicited by exteroceptive stimuli.
 
The parameters $T$, $h_{\rm int}$, and $h_{\rm ext}$ retain their meaning as contextual informativeness, learned valuation, and external information when defined with respect to the sensation or event under evaluation, allowing the same formal structure to be applied to visual, auditory, or somatosensory stimuli. 

Standardized affective stimulus sets such as GAPED~\cite{dan-glauser-2011} and OASIS~\cite{kurdi-2017}, together with the variability of valence--arousal relations observed across individuals and cultures~\cite{Yik:2023aa}, provide possible settings in which this broader applicability could be tested. At the same time, the association between  attention and affective experience reported in daily life~\cite{poerio-2024} suggests that interoception may retain a particularly important role without restricting the evaluative architecture to inner sensations. A detailed mapping between the model parameters and psychological or physiological measures remains an open empirical question.

\section{Outlook: dynamical extension}
\label{sec:outlook}

The present discussion has focused on the static organization of affect and on how different affective states emerge, compete, and reorganize within the sense-making landscape. The same framework also provides the ingredients needed to address their temporal evolution. Once we allow the model to evolve in time, the affective state can be represented as a trajectory in the valence--arousal plane,
$\bigl(\phi(t),A(t)\bigr)$, so that emotions may be approached as dynamical changes of affect.

At the minimal level, we can study the relaxation dynamics of $\phi$ on a fixed (static) landscape. In this case, the valence dynamics reads
\begin{equation}
\frac{d\phi}{dt}=-\Gamma_{\phi}\frac{\partial F}{\partial\phi},
\label{eq:kinetic}
\end{equation}
where $\Gamma_{\phi}>0$ is a kinetic coefficient setting the time scale of the evolution.
Writing $\phi(t)=\bar{\phi}+\delta\phi(t)$ and linearizing Eq.~(\ref{eq:kinetic}) with respect to $\delta\phi$ around a minimum, we find
\begin{equation}
\frac{d\,\delta\phi}{dt}\simeq -\Gamma_{\phi}
\left.\frac{\partial^2 F}{\partial\phi^2}\right|_{\phi=\bar{\phi}}
\,\delta\phi
=
-\frac{\Gamma_\phi}{\chi}\,\delta\phi,
\end{equation}
so that small deviations relax exponentially,
\begin{equation}
\delta\phi(t)=\delta\phi_{0}e^{-t/\tau_{\phi}},
\qquad
\tau_{\phi}=\frac{\chi}{\Gamma_{\phi}}.
\end{equation}
The linear relaxation of the valence is controlled by the same susceptibility that governs the static response of the system. This provides a direct link between the geometry of the static landscape and the time scale of affective adjustment. In particular, near the uncertainty regime of the symmetric case, where $\chi$ diverges, the relaxation time $\tau_\phi$ also diverges. The system therefore becomes increasingly sensitive to weak external perturbations and increasingly slow to return to the adapted state. This is the critical slowing down familiar from other complex systems, which has been proposed as an early-warning signal of transitions in mood\cite{vandeleemput-2014,wichers-2016} and is related to the emotional inertia observed in experience-sampling studies\cite{kuppens-2010}. The relation $\tau_\phi=\chi/\Gamma_\phi$ adds a specific, testable prediction: for a given sensation, susceptibility to externally supplied information and recovery time after a perturbation should covary.

The instantaneous position $\phi(t)$ also carries information about the evolving affective evaluation. As the system moves through the landscape, $F(\phi(t))$ specifies the instantaneous position of the current evaluation within the sense-making potential. Its possible relation to the transient activation of the affective state provides a natural direction for extending the present static definition of arousal. However, in the minimal setting considered here, i.e. the relaxation of $\phi(t)$ on a static landscape, arousal remain constant, being defined specifically through the depth of a stabilized minimum, $A(\bar{\phi})=-F(\bar{\phi})$ that doesn't evolve on a fixed landscape. 
The affective trajectory therefore describes, in this limit, the temporal evolution of valence at fixed arousal. A dynamical definition of arousal away from a stationary state consequently requires an extension of the present construction beyond the fixed-landscape approximation. 

A fuller dynamical theory should allow the various coefficients of the sense-making landscape to evolve in time. The resulting time-dependent landscape would then determine the evolution of the affective state.
Repeated experience may gradually modify the learned valuation encoded by $h_{\rm int}$, while changes in contextual sensory information and externally supplied information may be represented through time-dependent $T$ and $h_{\rm ext}$. On longer time scales, experience and adaptation may also modify the parameters $T_0$, $a_0$, and $b$ that characterize the individual landscape, allowing the framework to describe more persistent changes in affective organization. The positions and depths of the minima would then evolve in time, providing a genuinely time-dependent affective trajectory $\bigl(\bar{\phi}(t),A(t)\bigr)$ in the valence--arousal plane. Such an extension would make it possible to address relaxation, sensitization, habituation, feedback loops and learning within the same sense-making framework. 

The resulting description provides a natural mathematical platform for a dynamical theory of emotions. This perspective is closely related to Lewis' proposal~\cite{Lewis2005} that emotions should be understood as dynamical systems, potentially connecting emotion theory and neurobiology through processes of self-organization. While several dynamical approaches to emotion have been proposed (for a recent review see Ref.~[\onlinecite{Vanhasbroeck21}]), the present framework differs in taking the affective organization emerging from sense-making as the starting point for the dynamics. Within this view, emotions may, more specifically, be understood as dynamic attractors emerging from the temporal evolution of affect. This interpretation is consistent with Russell and Barrett's proposal that emotions are constructed upon affect~\cite{LFBarrettBook}, while naturally extending the enactive view of affect as an appraisal phenomenon grounded in sense-making~\cite{Colombetti:2019aa}. The present work establishes the static foundations of affect within the sense-making framework and opens a direct route toward a dynamical theory of affect and emotions to be developed in future works.

\section*{CONCLUSIONS}
\label{sec:conclusions}

Affect can be derived within the sense-making phase transition framework~\cite{Valenzuela_FrontPain2024, Fanfarillo_RSOS2026}, without introducing an additional phenomenological structure.
Each minimum of the sense-making potential defines an affective state: its position determines valence, while its depth determines arousal. Affect therefore emerges directly from the same landscape that organizes the evaluation of a sensation.

Three main results follow from this construction. First, valence intensity and arousal are distinct but generally not independent quantities, since both are determined by the same sense-making landscape. Second, a single valence dimension can simultaneously support stable pleasant and unpleasant states, allowing mixed affect to emerge from the multistability of the landscape without introducing independent positive- and negative-affect dimensions. Third, the evolution of the landscape itself generates the upper-open V-shaped relation between valence and arousal, whose symmetry is naturally modified by internal and external biases, while also organizing neutral, uncertain, sensitized, consolidated, and mixed-valence affective regimes.

The framework also opens a natural route toward affective dynamics. Allowing the landscape to evolve in time provides a basis for time-dependent valence and arousal and, more broadly, for understanding emotions as dynamic attractors emerging from the temporal evolution of affect. This possibility will be explored in future work.

The present framework remains directly connected to our previous phenomenological and microscopic work \cite{Valenzuela_FrontPain2024,Fanfarillo_RSOS2026}. This continuity is important: without introducing an independent model of affect, the present work reveals the affective interpretation already implicit in the same sense-making framework. 
At the microscopic level, affect emerges from the collective organization of a network of vital norms, represented by the spins in the Ising model\cite{Fanfarillo_RSOS2026}. At the macroscopic level, this organization is captured by the Ginzburg-Landau free-energy functional. 
Our model, linking microscopic interactions and macroscopic organization, provides then a coherent multiscale description grounded in condensed matter field theory\cite{Altland:2010aa}. In this regard, the predictions derived from the model provide a platform for guiding future theoretical and experimental research. More fundamentally, however, they will arise as consequences of a coherent mathematical structure grounded in the physics of collective phenomena.


\appendix
\section{Mathematical details of the sense-making landscape}
\label{app:landscape}

In this Appendix we provide the mathematical details underlying the landscape analysis presented in Sec.~\ref{sec:landscapes}. We derive the stationary solutions, their local stability, and the characteristic thresholds associated with the symmetric, internally biased, externally biased, and combined landscapes. We also derive the stationary-state expression for the arousal used in Sec.~\ref{sec:valence_arousal}.

Throughout the analysis we consider the local sense-making potential
\begin{equation}
F(\phi)
=
-h_{\rm ext}\phi
+\frac{a}{2}\phi^2
+\frac{h_{\rm int}}{3}\phi^3
+\frac{b}{4}\phi^4,
\label{eq:app_F}
\end{equation}
with
\begin{equation}
a=a_0(T-T_0),
\qquad
a_0>0,
\qquad
b>0.
\label{eq:app_a}
\end{equation}
The positive quartic coefficient $b$ ensures that the potential remains bounded from below within the quartic approximation considered here. If $b<0$, stability can instead be recovered by retaining higher-order terms. This possibility arises naturally in the microscopic framework of Ref.~[\onlinecite{Fanfarillo_RSOS2026}]. There, a Blume--Capel extension of the spin model introduces a local anisotropy that modifies the relative preference for neutral ($0$) and polarized ($\pm1$) evaluations in the network of interacting vital norms. In a suitable parameter regime, this anisotropy can drive the quartic coefficient negative, requiring the sense-making potential to be expanded beyond fourth order.

\subsection{General stationary-state, stability, and susceptibility conditions}
\label{app:general}

A stationary state $\bar{\phi}$ satisfies
\begin{equation}
\left.\frac{\partial F}{\partial\phi}\right|_{\bar{\phi}}
=
-h_{\rm ext}
+a\bar{\phi}
+h_{\rm int}\bar{\phi}^2
+b\bar{\phi}^3
=0.
\label{eq:app_stationary}
\end{equation}
Its local stability is determined by
\begin{equation}
\left.\frac{\partial^2F}{\partial\phi^2}\right|_{\bar{\phi}}
=
a+2h_{\rm int}\bar{\phi}
+3b\bar{\phi}^2.
\label{eq:app_curvature}
\end{equation}
Thus, $\partial^2F/\partial\phi^2>0$ identifies a locally stable minimum, whereas $\partial^2F/\partial\phi^2<0$ identifies an unstable stationary state. A change of local stability occurs when a stationary solution satisfies Eq.~\eqref{eq:app_stationary} while its curvature, Eq.~\eqref{eq:app_curvature}, vanishes.

For a locally stable minimum $\bar{\phi}$, the susceptibility to the external bias is obtained by differentiating Eq.~\eqref{eq:app_stationary} with respect to $h_{\rm ext}$. Since $\bar{\phi}$ depends implicitly on $h_{\rm ext}$,
\begin{equation}
-1+
\left(
a+2h_{\rm int}\bar{\phi}
+3b\bar{\phi}^2
\right)
\frac{\partial\bar{\phi}}{\partial h_{\rm ext}}
=0,
\end{equation}
and we can write
\begin{equation}
\chi(\bar{\phi})
\equiv
\frac{\partial\bar{\phi}}{\partial h_{\rm ext}}
=
\frac{1}{
a+2h_{\rm int}\bar{\phi}
+3b\bar{\phi}^2
}
=
\left[
\left.
\frac{\partial^2F}{\partial\phi^2}
\right|_{\bar{\phi}}
\right]^{-1}.
\label{eq:app_chi}
\end{equation}
The susceptibility is therefore the inverse local curvature of the landscape. In particular, it diverges when the curvature of a stationary branch vanishes, identifying a point at which the corresponding state is particularly sensitive to even weak external perturbations.

It is also useful to evaluate the potential directly along a stationary branch. Multiplying Eq.~\eqref{eq:app_stationary} by $\bar{\phi}$ gives
\begin{equation}
a\bar{\phi}^2
=
h_{\rm ext}\bar{\phi}
-h_{\rm int}\bar{\phi}^3
-b\bar{\phi}^4.
\label{eq:app_eliminate_a}
\end{equation}
Substitution into Eq.~\eqref{eq:app_F} yields
\begin{equation}
F(\bar{\phi})
=
-\frac{h_{\rm ext}}{2}\bar{\phi}
-\frac{h_{\rm int}}{6}\bar{\phi}^3
-\frac{b}{4}\bar{\phi}^4.
\label{eq:app_Fstationary}
\end{equation}
Using $A(\bar{\phi})=-F(\bar{\phi})$, one obtains
\begin{equation}
A(\bar{\phi})
=
\frac{h_{\rm ext}}{2}\bar{\phi}
+
\frac{h_{\rm int}}{6}\bar{\phi}^3
+
\frac{b}{4}\bar{\phi}^4,
\label{eq:app_Astationary}
\end{equation}
which is the stationary-state valence--arousal relation used in Eq.~\eqref{eq:emergent_VA}.

\subsection{Symmetric landscape}
\label{app:symmetric}

For $h_{\rm int}=h_{\rm ext}=0$, the stationary condition becomes
\begin{equation}
\left.
\frac{\partial F}{\partial\phi}
\right|_{\bar{\phi}}
=
\bar{\phi}
\left(
a+b\bar{\phi}^2
\right)
=0.
\end{equation}
The corresponding stationary solutions are
\begin{equation}
\bar{\phi}_0=0,
\qquad
\bar{\phi}_{\pm}
=
\pm\sqrt{-\frac{a}{b}}.
\label{eq:app_symm_roots}
\end{equation}
The neutral solution $\bar{\phi}_0$ exists for all values of $T$. The two nonzero solutions $\bar{\phi}_{\pm}$ are real only when the argument of the square root is non-negative. Since $a=a_0(T-T_0)$, with $a_0>0$ and $b>0$, the polarized stationary branches therefore appear at $T=T_0$ and become distinct from the neutral solution for $T<T_0$.

The local stability of these stationary solutions follows from the curvature of the potential. For the neutral branch,
\begin{equation}
\left.
\frac{\partial^2F}{\partial\phi^2}
\right|_{\bar{\phi}_0}
=
a.
\end{equation}
Thus, $\bar{\phi}_0$ is stable for $T>T_0$, becomes marginal at $T=T_0$, where its curvature vanishes, and is unstable for $T<T_0$. Along the stable neutral branch, the susceptibility behaves as $\chi(\bar{\phi}_0)=1/a$ and therefore diverges as $T\rightarrow T_0^{+}$.

For the two nonzero solutions, Eq.~\eqref{eq:app_symm_roots} gives
\begin{equation}
\left.
\frac{\partial^2F}{\partial\phi^2}
\right|_{\bar{\phi}_{\pm}}
=
-2a>0,
\end{equation}
so that both polarized branches are stable for $T<T_0$. Their susceptibility is $\chi(\bar{\phi}_{\pm})=-1/(2a)$ and diverges as $T\rightarrow T_0^{-}$. Thus, at $T_0$ the neutral branch loses local stability while two stable polarized branches, corresponding to the pleasant and unpleasant states, emerge continuously from it, producing the pitchfork bifurcation discussed in the main text.

The neutral state has zero valence and zero arousal. The pleasant and unpleasant branches instead acquire finite valence according to Eq.~\eqref{eq:app_symm_roots}, while their arousal is
\begin{equation}
A(\bar{\phi}_{\pm})
=
-F(\bar{\phi}_{\pm})
=
\frac{a^2}{4b}.
\end{equation}

\subsection{Internal bias: $h_{\rm int}\neq0$, $h_{\rm ext}=0$}
\label{app:hint}

For a finite internal bias the stationary condition factorizes as
\begin{equation}
\left.\frac{\partial F}{\partial\phi}\right|_{\bar{\phi}} = \bar{\phi}
\left(
a+h_{\rm int}\bar{\phi}
+b\bar{\phi}^2
\right)=0.
\label{eq:app_hint_stationary}
\end{equation}
The neutral branch $\bar{\phi}_0=0$ therefore remains stationary for all $T$. The two additional branches are
\begin{equation}
\bar{\phi}_{\pm}
=
\frac{
-h_{\rm int}
\pm
\sqrt{h_{\rm int}^2-4ab}
}{2b}.
\label{eq:app_hint_roots}
\end{equation}
We derive below the characteristic thresholds $T_2$, $T_1$, $T_c$, and $T_0$. For definiteness we take $h_{\rm int}>0$; reversing its sign exchanges the pleasant and unpleasant sectors without changing the values of the thresholds.

\subsubsection{Inflection threshold $T_2$}

The points where the curvature vanishes are obtained from
\begin{equation}
\frac{\partial^2F}{\partial\phi^2}
=
a+2h_{\rm int}\phi+3b\phi^2
=0,
\end{equation}
which gives
\begin{equation}
\phi_{\rm inf}^{\pm}
=
\frac{-h_{\rm int}\pm
\sqrt{h_{\rm int}^2-3ab}}{3b}.
\label{eq:app_inflection_hint}
\end{equation}
Real solutions first appear when the discriminant of
this quadratic vanishes,
\begin{equation}
h_{\rm int}^2-3ab=0.
\end{equation}
This gives
\begin{equation}
a(T_2) = \frac{h_{int}^2}{3b}
\end{equation}
or 
\begin{equation}
T_2
=
T_0
+
\frac{1}{3}
\frac{h_{\rm int}^2}{a_0b}.
\label{eq:app_T2}
\end{equation}
where we used that $a=a_0(T-T_0)$. 
At $T=T_2$, the two curvature-zero solutions $\phi_{\rm inf}^{\pm}$ coincide, while for $T<T_2$ they separate into two inflection points of the landscape. Importantly, these are not stationary states. Thus, $T_2$ marks a geometrical reorganization of the landscape, not a change in the local stability of an affective state. 

\subsubsection{Bifurcation threshold $T_1$}

The two nonzero stationary solutions in Eq.~\eqref{eq:app_hint_roots} first appear when their discriminant vanishes, $h_{\rm int}^2-4ab=0$.
This gives $a(T_1) = h_{\rm int}^2 / 4b$, i.e.
\begin{equation}
T_1
=
T_0
+
\frac{1}{4}
\frac{h_{\rm int}^2}{a_0b}.
\label{eq:app_T1}
\end{equation}
At $T_1$ the two branches coincide 
\begin{equation}
\bar{\phi}_{+}
=
\bar{\phi}_{-}
=
-\frac{h_{\rm int}}{2b}.
\label{eq:app_phi_T1}
\end{equation}
and the curvature vanishes as follow from Eq.~\eqref{eq:app_curvature}. Thus, the susceptibility of the emerging branch diverges at $T_1$.

For $T<T_1$, the two branches separate. Substitution of the solutions in Eq.~\eqref{eq:app_curvature} shows that, for $h_{\rm int}>0$, the more negative branch $\bar{\phi}_{-}$ is a stable local minimum, which we identify with $\bar{\phi}_{unpl}$. The branch $\bar{\phi}_{+}$ is instead unstable for $T_0<T<T_1$ and becomes the stable pleasant minimum only below $T_0$.

\subsubsection{Coexistence threshold $T_c$}

The coexistence threshold $T_c$ is defined by equality in depth between the neutral and unpleasant minima. Since $F(\bar{\phi}_0)=F(0)=0 $, the coexistence condition is simply
\begin{equation}
F(\bar{\phi}_{unpl})=0.
\label{eq:app_coexistence_hint}
\end{equation}
For $h_{\rm int}>0$, the unpleasant stationary branch is
\begin{equation}
\bar{\phi}_{unpl}
=
\bar{\phi}_{-}
=
\frac{-h_{\rm int}
-\sqrt{h_{\rm int}^2-4ab}}{2b}.
\end{equation}
Substituting this expression into the sense-making potential and imposing Eq.~\eqref{eq:app_coexistence_hint}, one find 
\begin{equation}
a(T_c)
=
\frac{2}{9}
\frac{h_{\rm int}^2}{b}.
\end{equation}
Using $a=a_0(T-T_0)$ therefore gives
\begin{equation}
T_c
=
T_0
+
\frac{2}{9}
\frac{h_{\rm int}^2}{a_0b}.
\label{eq:app_Tc}
\end{equation}
For $T_c<T<T_1$, the unpleasant minimum is locally stable but lies higher in the sense-making potential than the neutral minimum, corresponding to the latent unpleasant state introduced in the main text. At $T=T_c$, the two minima become degenerate, with $F=0$ and therefore the same vanishing arousal. Their local curvatures remain finite, so $T_c$ is not a local-stability threshold but a coexistence threshold at which global stability is exchanged. For $T<T_c$, the unpleasant minimum becomes the globally adapted state.

The complete ordering is
\begin{equation}
T_2>T_1>T_c>T_0.
\label{eq:app_hint_ordering}
\end{equation}
All three additional thresholds are displaced above $T_0$ by an amount proportional to $h_{\rm int}^2/a_0b$,
which explains why their position depends on the magnitude, but not the sign, of the internal bias.

\subsubsection{Loss of stability at $T_0$ and biased mixed valence}

The curvature of the neutral state is
\begin{equation}
\left.
\frac{\partial^2F}{\partial\phi^2}
\right|_{\bar{\phi}_0}
=a.
\end{equation}
It therefore vanishes at $T=T_0$, where the neutral state becomes marginal, and changes sign for $T<T_0$, where the neutral state becomes unstable.
From Eq.~\eqref{eq:app_hint_roots}, $\bar{\phi}_{+}\rightarrow0$ as
$T\rightarrow T_0$, so that this branch merges with the neutral solution at $T_0$. For $T<T_0$, $\bar{\phi}_{+}$ moves into the positive-valence sector and becomes the stable pleasant minimum $\bar{\phi}_{pl}$, while $\bar{\phi}_{-}=\bar{\phi}_{unpl}$ remains the globally stable unpleasant minimum. This produces the biased mixed-valence landscape discussed in the main text.

\subsection{External bias: $h_{\rm ext}\neq0$, $h_{\rm int}=0$}
\label{app:hext}

For a finite external bias the stationary condition becomes
\begin{equation}
\left.
\frac{\partial F}{\partial\phi}
\right|_{\bar{\phi}}
=
-h_{\rm ext}
+a\bar{\phi}
+b\bar{\phi}^3
=0.
\label{eq:app_hext_stationary}
\end{equation}
We consider $h_{\rm ext}>0$ for definiteness; the case $h_{\rm ext}<0$ follows by exchanging the pleasant and unpleasant sectors. Unlike the internally biased case, $\bar{\phi}=0$ is not a stationary solution. The external bias therefore selects the positive-valence sector at any finite $T$.

At high $T$, where $a$ is large and positive, the cubic contribution in Eq.~\eqref{eq:app_hext_stationary} is subleading and
\begin{equation}
\bar{\phi}_{pl}
\simeq
\frac{h_{\rm ext}}{a}.
\end{equation}
Using Eq.~\eqref{eq:app_Astationary}, the corresponding arousal is
\begin{equation}
A(\bar{\phi}_{pl})
\simeq
\frac{h_{\rm ext}^2}{2a}.
\end{equation}
Both valence and arousal tend continuously to zero in the limit $T\rightarrow\infty$.

At $T=T_0$, where $a=0$, the stationary equation gives
\begin{equation}
\bar{\phi}_{pl}(T_0)
=
\left(
\frac{h_{\rm ext}}{b}
\right)^{1/3}.
\label{eq:app_hext_phiT0}
\end{equation}
and the curvature of this stationary state is
\begin{equation}
\left.
\frac{\partial^2F}{\partial\phi^2}
\right|_{\bar{\phi}_{pl}}
=
3b\bar{\phi}_{pl}^2>0,
\end{equation}
so the adapted branch remains locally stable. For finite $h_{\rm ext}$, $T_0$ is therefore neither a loss-of-stability threshold nor a point of divergent susceptibility.

Its geometrical significance can instead be understood by examining the curvature of the full landscape. From the general expression in Eq.~\eqref{eq:app_curvature}, setting $h_{\rm int}=0$ gives
\begin{equation}
\frac{\partial^2F}{\partial\phi^2}
=
a+3b\phi^2.
\end{equation}
For $a>0$ the curvature is positive for all $\phi$, whereas for $a<0$ two points of vanishing curvature appear,
\begin{equation}
\phi_{\rm inf}^{\pm}
=
\pm\sqrt{-\frac{a}{3b}},
\end{equation}
corresponding to two inflection points of the landscape. Thus, in the externally biased case, crossing $T_0$ marks a geometrical reorganization of the potential rather than a change in the local stability of the adapted state.

\subsubsection{Appearance of the opposite-valence minimum}

A second minimum appears when a stationary solution simultaneously develops zero curvature. This is the spinodal condition that corresponds to the limit of local metastability, where a local minimum and the adjacent maximum merge into a stationary point of vanishing curvature satisfying:
\begin{equation}
\left.
\frac{\partial F}{\partial\phi}
\right|_{\bar{\phi}_s}
=
-h_{\rm ext}
+a\bar{\phi}_s
+b\bar{\phi}_s^3
=0,
\end{equation}
together with
\begin{equation}
\left.
\frac{\partial^2F}{\partial\phi^2}
\right|_{\bar{\phi}_s}
=
a+3b\bar{\phi}_s^2
=0.
\label{eq:app_hext_spinodal}
\end{equation}
Solving these two equations gives
\begin{equation}
\bar{\phi}_s
=
-\left(
\frac{h_{\rm ext}}{2b}
\right)^{1/3},
\label{eq:app_phi_s}
\end{equation}
and
\begin{equation}
a(T_s)
=
-\frac{3}{2^{2/3}}
\left(
b h_{\rm ext}^2
\right)^{1/3}.
\end{equation}
Using $a=a_0(T-T_0)$, we obtain
\begin{equation}
T_s
=
T_0
-
\frac{3}{2^{2/3}a_0}
\left(
b h_{\rm ext}^2
\right)^{1/3}.
\label{eq:app_Ts}
\end{equation}
At $T_s$ the curvature of the emerging stationary branch vanishes and its susceptibility diverges. For $T<T_s$, the two newly separated stationary branches correspond to a local minimum in the negative-valence sector, $\bar{\phi}_{unpl}$, and an unstable maximum. The positive-valence minimum $\bar{\phi}_{pl}$ remains globally stable.

\subsubsection{Onset of positive arousal of the latent state}

The newly opposite-valence state emerging at $T_s$ does not yet have positive arousal. Indeed, substituting Eq.~\eqref{eq:app_phi_s} into
Eq.~\eqref{eq:app_Fstationary} shows that
\begin{equation}
F(\bar{\phi}_s)>F(0)=0.
\end{equation}
The emerging minimum therefore lies above the reference level and has not yet acquired positive arousal.

Upon further lowering $T$, the latent unpleasant minimum deepens and eventually crosses the reference level $F=0$. Solving
\begin{equation}
F(\bar{\phi}_{unpl})=0
\end{equation}
together with the stationary condition gives
\begin{equation}
a
=
-\frac{3}{2^{1/3}}
\left(
b h_{\rm ext}^2
\right)^{1/3}.
\end{equation}
This condition is fulfilled only at a temperature below $T_s$. The latent unpleasant minimum therefore first appears at $T_s$ and acquires positive arousal only upon further lowering $T$.

\subsection{Combined internal and external bias}
\label{app:combined}

We finally consider the case in which both biases are finite,
$h_{\rm int}\neq0$ and $h_{\rm ext}\neq0$. The stationary states satisfy
\begin{equation}
\left.
\frac{\partial F}{\partial\phi}
\right|_{\bar{\phi}}
=
-h_{\rm ext}
+a\bar{\phi}
+h_{\rm int}\bar{\phi}^2
+b\bar{\phi}^3
=0.
\label{eq:app_combined_stationary}
\end{equation}
As in the externally biased case, $\bar{\phi}=0$ is no longer a stationary
solution. The main effect of combining the two biases is therefore not the
appearance of a new type of stationary condition, but the modification of
the characteristic thresholds derived above.

\subsubsection{Inflection and bifurcation thresholds}

The external bias $h_{\rm ext}$ does not enter the curvature,
\begin{equation}
\frac{\partial^2F}{\partial\phi^2}
=
a+2h_{\rm int}\phi+3b\phi^2.
\end{equation}
The condition for the first appearance of inflection points is therefore unchanged from the internally biased case, and
\begin{equation}
T_2
=
T_0
+
\frac{1}{3}
\frac{h_{\rm int}^2}{a_0b}.
\end{equation}

The bifurcation threshold $T_1$, instead, is shifted by the external bias. As in the previous cases, it is determined by requiring a stationary solution, Eq.~\eqref{eq:app_stationary}, to have vanishing curvature, Eq.~\eqref{eq:app_curvature}. Combining these two conditions gives
\begin{equation}
2b\bar{\phi}_1^3
+
h_{\rm int}\bar{\phi}_1^2
+
h_{\rm ext}
=0,
\label{eq:app_combined_phi1}
\end{equation}
while the corresponding threshold is
\begin{equation}
T_1
=
T_0
-
\frac{
2h_{\rm int}\bar{\phi}_1}{a_0}
- \frac{3b\bar{\phi}_1^2}{a_0}.
\label{eq:app_combined_T1}
\end{equation}
%
%
For $h_{ext}\rightarrow 0$, these expressions recover the internally
biased result:
\begin{equation}
\bar{\phi}_1
\rightarrow
-\frac{h_{int}}{2b},
\end{equation}
as given in Eq.~\eqref{eq:app_phi_T1}, and
\begin{equation}
T_1
\rightarrow
T_0+\frac{1}{4}\frac{h_{int}^2}{a_0b},
\end{equation}
as given in Eq.~\eqref{eq:app_T1}.

The effect of a weak external bias can be made explicit by expanding around this solution. To leading order in $h_{ext}$, the bifurcation threshold becomes
\begin{equation}
T_1(h_{int},h_{ext})
=
T_0
+\frac{1}{4}\frac{h_{int}^2}{a_0b}
-\frac{2b}{a_0}\frac{h_{ext}}{h_{int}}
+\mathcal{O}(h_{ext}^2).
\label{eq:T1_external}
\end{equation}
Thus, unlike $T_2$, the bifurcation threshold is shifted already at linear order by the external bias. With the sign convention used here, equal signs of $h_{int}$ and $h_{ext}$ correspond to competing tendencies and lower $T_1$, whereas opposite signs correspond to
reinforcing tendencies and shift $T_1$ upward.

\subsubsection{Coexistence threshold and arousal}

The coexistence threshold $T_c$ is defined by the presence of two locally stable stationary states, $\bar{\phi}_{c,+}$ and $\bar{\phi}_{c,-}$, with equal depth,
\begin{equation}
F(\bar{\phi}_{c,+})
=
F(\bar{\phi}_{c,-}).
\label{eq:app_combined_coexistence}
\end{equation}
At $T_c$, each of these states also satisfies the stationary condition, Eq.~\eqref{eq:app_combined_stationary}. Solving the latter for $a=a_0(T_c-T_0)$ gives
\begin{equation}
T_c
=
T_0
+
\frac{1}{a_0}
\left[
\frac{h_{\rm ext}}{\bar{\phi}_{c,\pm}}
-h_{\rm int}\bar{\phi}_{c,\pm}
-b\bar{\phi}_{c,\pm}^2
\right].
\label{eq:app_Tc_param}
\end{equation}
\noindent
which is an analogous result to Eq.~(\ref{eq:app_combined_T1}), this time for the coexistence threshold.

An explicit expression for the coexistence temperature, namely,
\begin{equation}
T_c(h_{\rm int},h_{\rm ext})
=
T_0
+
\frac{2}{9}
\frac{h_{\rm int}^2}{a_0b}
-
\frac{3b}{a_0}
\frac{h_{\rm ext}}{h_{\rm int}},
\label{eq:app_Tc_combined}
\end{equation}
is obtained directly by introducing a change of variable $\phi \to \psi -h_{int}/(3b)$ in the sense-making potential, Eq.~(\ref{eq:model}), which removes the third-order term, and by imposing a null linear term, so that parity of the potential makes possible the above equal-depth condition, Eq.~(\ref{eq:app_combined_coexistence}). For $h_{\rm ext}=0$, Eq.~(\ref{eq:app_Tc_combined}) reduces to the coexistence threshold of the internally biased landscape, Eq.~\eqref{eq:app_Tc}. A finite external bias therefore shifts the contextual condition at which the two minima exchange global stability. 

At the coexistence temperature, $T=T_c$, stationarity (now $\partial F(\psi)/\partial \psi |_{\bar{\psi}_c}=0$) yields the two coexisting minima:
\begin{equation}
\bar{\phi}_{c,\pm}
=
-\frac{h_{\rm int}}{3b}
\pm
\sqrt{
\frac{h_{\rm int}^2}{9b^2}
+
3\frac{h_{\rm ext}}{h_{\rm int}}
}.
\label{eq:app_phi_c_combined}
\end{equation}

With the sign convention used here, $h_{\rm int}>0$
favors negative valence whereas $h_{\rm ext}>0$ favors positive valence.
Consequently, equal signs of $h_{\rm int}$ and $h_{\rm ext}$ correspond to
competing tendencies, whereas opposite signs reinforce the same valence
sector.

The external bias also changes the nature of the two states that coexist at
$T_c$. To determine whether they belong to opposite or to the same valence
sector, it is useful to consider the product of the two solutions in
Eq.~\eqref{eq:app_phi_c_combined},
\begin{equation}
\bar{\phi}_{c,+}\bar{\phi}_{c,-}
=
-3\frac{h_{\rm ext}}{h_{\rm int}}.
\label{eq:app_product_phi_c}
\end{equation}
If $h_{\rm ext}/h_{\rm int}>0$, the product is negative and the two
coexisting minima necessarily have opposite valence. If
$h_{\rm ext}/h_{\rm int}<0$, the product is positive and, when the two
minima exist, they belong to the same valence sector but correspond to
different valence intensities.

The coexistence condition also determines the arousal of the two competing states. Since the two minima have equal depth at $T_c$, they necessarily have the same arousal, given by
\begin{equation}
A_c
=
\frac{9b}{4}
\frac{h_{\rm ext}^2}{h_{\rm int}^2}.
\label{eq:app_Ac}
\end{equation}
For $h_{\rm ext}=0$, $A_c=0$ and the internally biased result is recovered. A finite external bias therefore not only shifts the coexistence threshold and displaces the competing states in valence, but also makes their common arousal at coexistence finite.

\subsubsection{Spinodals and asymmetric hysteresis}

Finally, the combined landscape allows us to determine the limits of
metastability when $h_{\rm ext}$ is varied at fixed $T$. A local minimum
disappears when it simultaneously satisfies the stationary condition and
has vanishing curvature. The corresponding positions are
\begin{equation}
\bar{\phi}_{s,\pm}
=
\frac{
-h_{\rm int}
\pm
\sqrt{h_{\rm int}^2-3ab}
}{3b},
\label{eq:app_spinodal_phi_combined}
\end{equation}
provided $h_{\rm int}^2-3ab\geq0$. The corresponding external biases are
\begin{equation}
h_{{\rm ext},s}^{\pm}
=
-\bar{\phi}_{s,\pm}^2
\left(
h_{\rm int}
+2b\bar{\phi}_{s,\pm}
\right).
\label{eq:app_spinodal_hext}
\end{equation}
For $h_{\rm int}=0$, the two spinodal fields are symmetric under
$h_{\rm ext}\rightarrow-h_{\rm ext}$. A finite internal bias breaks this
symmetry, so that in general
\begin{equation}
\left|
h_{{\rm ext},s}^{+}
\right|
\neq
\left|
h_{{\rm ext},s}^{-}
\right|.
\end{equation}
Thus, if the system remains on a metastable branch as $h_{\rm ext}$ is
varied, the external bias required to destabilize an affective state can
differ in the two directions. This provides the static landscape basis for
the asymmetric hysteretic response discussed in Sec.~\ref{sec:landscapes};
its actual realization additionally depends on the dynamics of the system.

\bibliography{Landauaffect}

\end{document}